AI Ethics and Governance in Practice Programme

# AI Ethics and Governance in Practice

## An Introduction

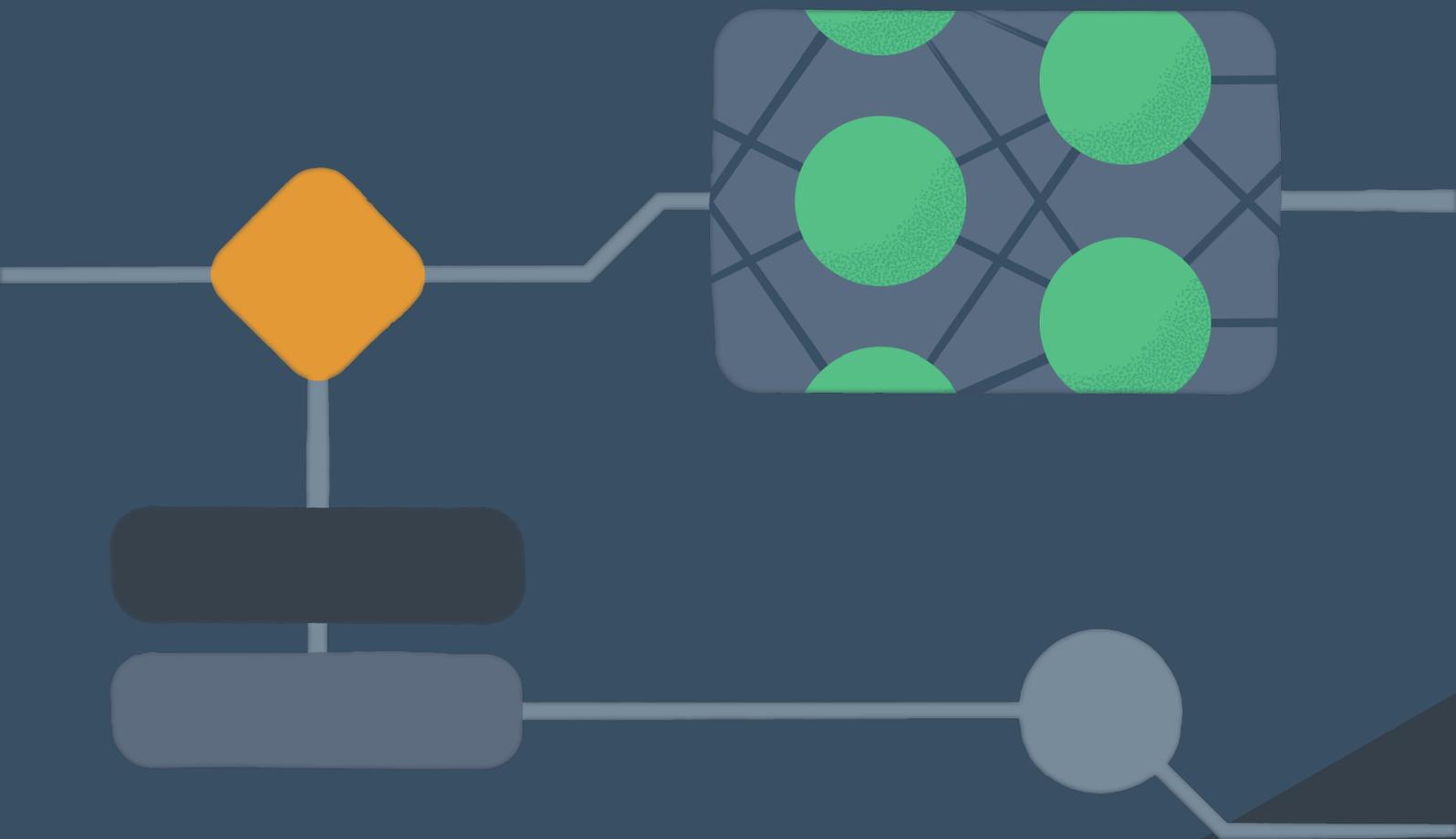

**For Facilitators**
This workbook is annotated to support facilitators in delivering the accompanying activities.

The Alan Turing Institute

# Acknowledgements

This workbook was written by David Leslie, Cami Rincón, Morgan Briggs, Antonella Perini, Smera Jayadeva, Ann Borda, SJ Bennett, Christopher Burr, Mhairi Aitken, Michael Katell, and Claudia Fischer.

The creation of this workbook would not have been possible without the support and efforts of various partners and collaborators. As ever, all members of our brilliant team of researchers in the Ethics Theme of the Public Policy Programme at The Alan Turing Institute have been crucial and inimitable supports of this project from its inception several years ago, as have our Public Policy Programme Co-Directors, Helen Margetts and Cosmina Dorobantu. We are deeply thankful to Conor Rigby, who led the design of this workbook and provided extraordinary feedback across its iterations. We also want to acknowledge Johnny Lighthands, who created various illustrations for this document, and Alex Krook and John Gilbert, whose input and insights helped get the workbook over the finish line. Special thanks must be given to the Equality and Human Rights Commission for helping us test the activities and review the content included in this workbook. Lastly, we want to thank Semeli Hadjiloizou (The Alan Turing Institute) for her meticulous peer review and timely feedback, which greatly enriched this document.

This work was supported by Wave 1 of The UKRI Strategic Priorities Fund under the EPSRC Grant EP/W006022/1, particularly the Public Policy Programme theme within that grant & The Alan Turing Institute; Towards Turing 2.0 under the EPSRC Grant EP/W037211/1 & The Alan Turing Institute; and the Ecosystem Leadership Award under the EPSRC Grant EP/X03870X/1 & The Alan Turing Institute.

Cite this work as: Leslie, D., Rincón, C., Briggs, M., Perini, A., Jayadeva, S., Borda, A., Bennett, SJ. Burr, C., Aitken, M., Katell, M., Fischer, C. (2023). *AI Ethics and Governance in Practice: An Introduction.* The Alan Turing Institute.

# Contents

# About the AI Ethics and Governance in Practice Workbook Series

## Who We Are

The Public Policy Programme at The Alan Turing Institute was set up in May 2018 with the aim of developing research, tools, and techniques that help governments innovate with data-intensive technologies and improve the quality of people's lives. We work alongside policymakers to explore how data science and artificial intelligence can inform public policy and improve the provision of public services. We believe that governments can reap the benefits of these technologies only if they make considerations of ethics and safety a first priority.

## Origins of the Workbook Series

In 2019, The Alan Turing Institute's Public Policy Programme, in collaboration with the UK's Office for Artificial Intelligence and the Government Digital Service, published the [UK Government's official Public Sector Guidance on AI Ethics and Safety](). This document provides end-to-end guidance on how to apply principles of AI ethics and safety to the design, development, and implementation of algorithmic systems in the public sector. It provides a governance framework designed to assist AI project teams in ensuring that the AI technologies they build, procure, or use are ethical, safe, and responsible.

In 2021, the UK's National AI Strategy recommended as a 'key action' the update and expansion of this original guidance. From 2021 to 2023, with the support of funding from the Office for AI and the Engineering and Physical Sciences Research Council as well as with the assistance of several public sector bodies, we undertook this updating and expansion. The result is the AI Ethics and Governance in Practice Programme, a bespoke series of eight workbooks and a forthcoming digital platform designed to equip the public sector with tools, training, and support for adopting what we call a Process-Based Governance (PBG) Framework to carry out projects in line with state-of-the-art practices in responsible and trustworthy AI innovation.

# About the Workbooks

The AI Ethics and Governance in Practice Programme curriculum is composed of a series of eight workbooks. Each of the workbooks in the series covers how to implement a key component of the PBG Framework. These include Sustainability, Technical Safety, Accountability, Fairness, Explainability, and Data Stewardship. Each of the workbooks also focuses on a specific domain, so that case studies can be used to promote ethical reflection and animate the Key Concepts.

**Programme Curriculum: AI Ethics and Governance in Practice Workbook Series**

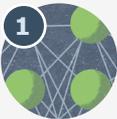
**1 AI Ethics and Governance in Practice: An Introduction**
*Multiple Domains*

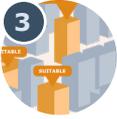
**5 Responsible Data Stewardship in Practice**
*AI in Policing and Criminal Justice*

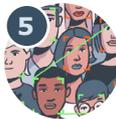
**2 AI Sustainability in Practice Part One**
*AI in Urban Planning*

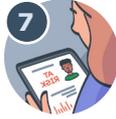
**6 AI Safety in Practice**
*AI in Transport*

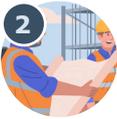
**3 AI Sustainability in Practice Part Two**
*AI in Urban Planning*

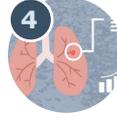
**7 AI Transparency and Explainability in Practice**
*AI in Social Care*

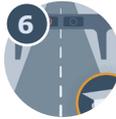
**4 AI Fairness in Practice**
*AI in Healthcare*

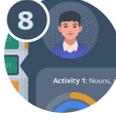
**8 AI Accountability in Practice**
*AI in Education*

Taken together, the workbooks are intended to provide public sector bodies with the skills required for putting AI ethics and governance principles into practice through the full implementation of the guidance. To this end, they contain activities with instructions for either facilitating or participating in capacity-building workshops.

Please note, these workbooks are living documents that will evolve and improve with input from users, affected stakeholders, and interested parties. We need your participation. Please share feedback with us at policy@turing.ac.uk.

## Programme Roadmap

The graphic below visualises this workbook in context alongside key frameworks, values and principles discussed within this programme. For more information on how these elements build upon one another, visit the Part Two: The Sociotechnical Aspect of the AI/ML Project Lifecycle section of this workbook.

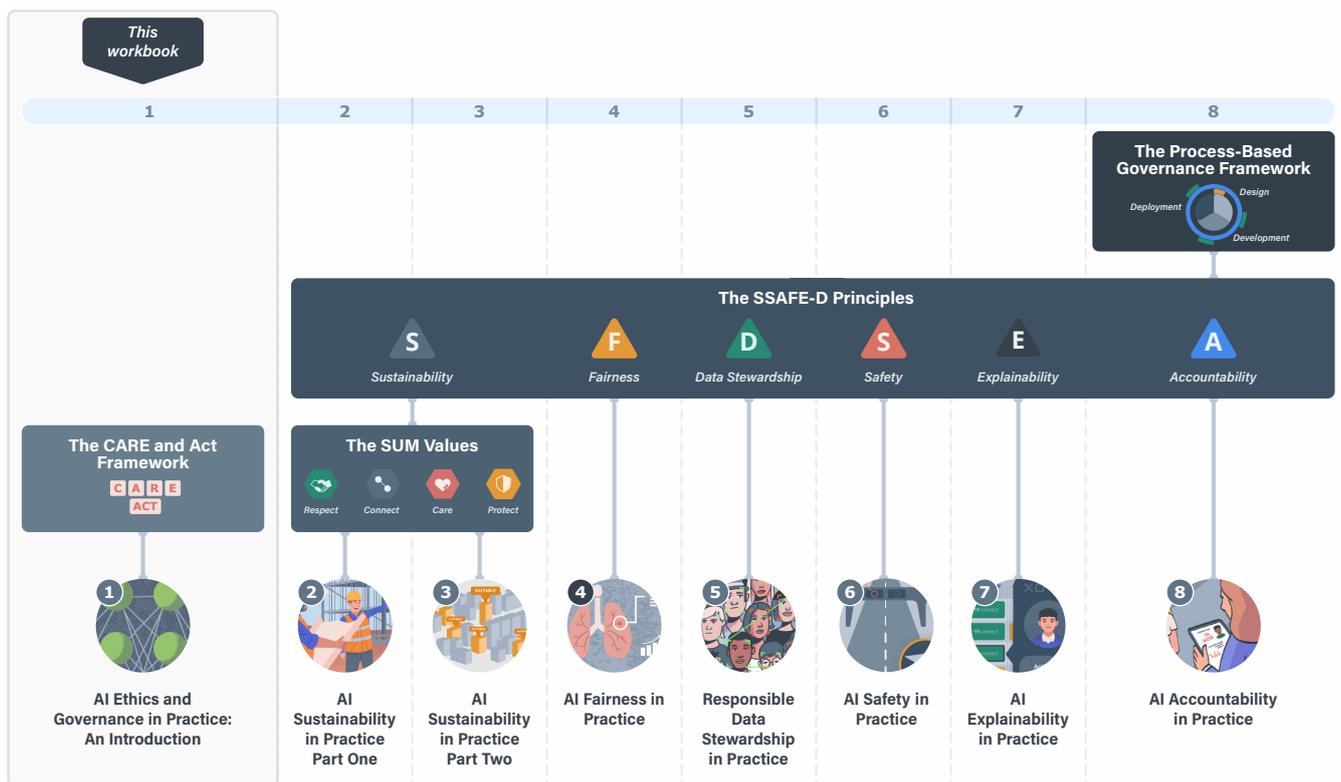

# Intended Audience

The workbooks are primarily aimed at civil servants engaging in the AI Ethics and Governance in Practice Programme as either AI Ethics Champions delivering the curriculum within their organisations by facilitating peer-learning workshops, or participants completing the programme by attending workshops. Anyone interested in learning about AI ethics, however, can make use of the programme curriculum, the workbooks, and resources provided. These have been designed to serve as stand-alone, open access resources. Find out more at turing.ac.uk/ai-ethics-governance.

There are two versions of each workbook:

- **Annotated workbooks** (such as this document) are intended for facilitators. These contain guidance and resources for preparing and facilitating training workshops.

- **Non-annotated workbooks** are intended for workshop participants to engage with in preparation for, and during, workshops.

# Introduction to This Workbook

The purpose of this workbook is to provide an explanatory introduction to artificial intelligence (AI) and machine learning (ML). The workbook is divided into two sections, Key Concepts and Activities. The first section provides an introduction to some key technical, sociotechnical, and ethical concepts that offer a knowledge base for those engaging in the AI Ethics and Governance in Practice Programme. Material covered in the first section includes a technical introduction to AI and ML, followed by an overview of the AI Ethics and Governance frameworks that will be explored in more detail in subsequent workbooks. The second section provides a set of group-based activities that are intended to bring the ideas explored in the first section to life. Here are the two sections at a glance:

**Key Concepts Section**

This section provides content for workshop participants and facilitators to engage with prior to attending each workshop. The Key Concepts demystify AI and ML by discussing foundational components that make up AI systems, providing definitions of key terms, an overview of the stages for building AI models, and a brief introduction to AI ethics. Topics discussed include:

**Part One: Introduction to Artificial Intelligence**

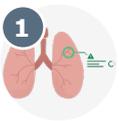
Introduction to Artificial Intelligence (AI)

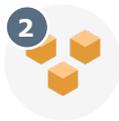
Technical Components of AI and Machine Learning (ML)

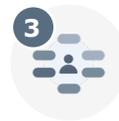
Examples of the Use of AI Systems in the Public Sector

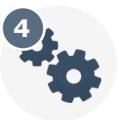
Types of ML

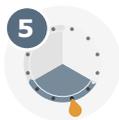
Stages of the AI/ML Project Lifecycle

**Part Two: The Sociotechnical Aspect of the AI/ML Project Lifecycle**

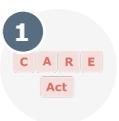
The CARE and Act Framework

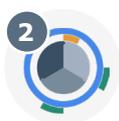
AI Ethics and Governance in Practice

## Activities Section

This section contains instructions for group-based activities (each corresponding to a section in the Key Concepts). These activities are intended to increase understanding of Key Concepts by using them.

*Case studies within the AI Ethics and Governance in Practice workbook series are grounded in public sector use cases, but do not reference specific AI projects.*

**Collective Image**

Illustrate the team's collective image of AI.

**Exploring Public Sector AI**

Consider public sector AI and participants' potential status as data subjects.

**Mapping the AI/ML Project Lifecycle**

Develop an understanding of the different steps taken in designing, developing, and deploying AI systems.

**Exploring the CARE and Act Framework**

Understand how the CARE and Act Framework can serve as a starting point for establishing habits of critical reflection and responsible innovation across the AI/ML project lifecycle.

**Revisiting Collective Image**

Revisit the team's collective image of AI.

> **Note for Facilitators**
>
> Additionally, you will find facilitator instructions (and where appropriate, considerations) required for facilitating activities and delivering capacity-building workshops.

AI Ethics and Governance in Practice: An Introduction

# Key Concepts

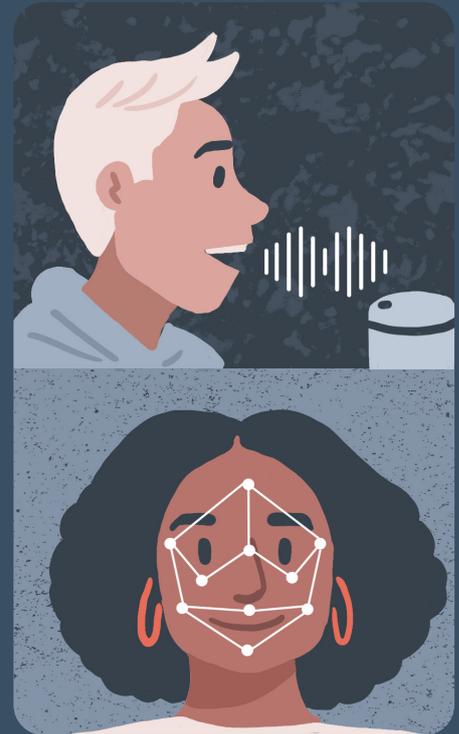

# Part One: Introduction to Artificial Intelligence (AI)

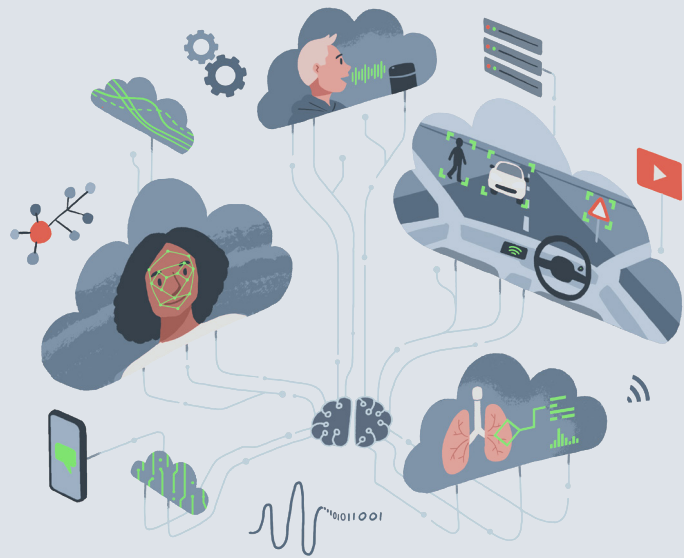

There are many ways that Artificial Intelligence (AI) has been defined over the last several decades, but for the purposes of this workbook we will stick to defining it by describing what it does, i.e. what role it plays in the human world: **"AI system" means any computational system (or a combination of such systems) that uses methods derived from statistics or other mathematical techniques to carry out tasks that are commonly associated with, or would otherwise require, human intelligence and that either assists or replaces the judgment of human decision-makers in carrying out those tasks.**

Such tasks include (but are not limited to) prediction, planning, classification, pattern recognition, organisation, perception, speech/sound/image recognition, text/sound/image generation, language translation, communication, learning, representation, and problem-solving.

In an increasingly digitally interconnected world, advancements in AI research and innovation have the potential to benefit society across many domains, from transport and education to climate science, biomedical research, health, and social care. However, in order to realise these benefits, AI needs to be carefully designed and implemented in a way which ensures its ethical permissibility and accounts for its limitations. This series of workbooks aims to facilitate the responsible and trustworthy development of AI systems which are suited to their contexts of deployment.

**Examples of AI Systems**

AI-supported healthcare has helped clinicians to spot early signs of illness and diagnose diseases.[1]

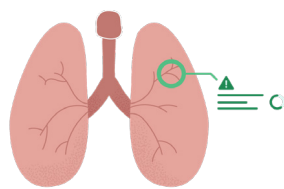

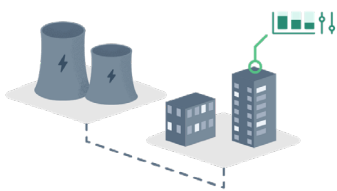

In energy and utilities management, AI systems have been used to predict energy consumption across electric grids, reducing unnecessary energy generation.[2]

By bringing together diverse groups from around the globe through real-time speech-to-speech translation, AI systems are enabling humans to successfully confront an ever-widening range of societal challenges.[3]

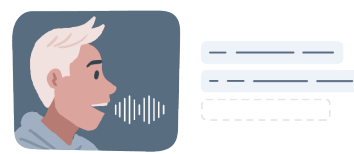

These successes indicate the potential of AI technologies to aid in addressing societal challenges. Furthermore, as data access and computational power increase, these tools will become more powerful and capable, playing an increasing role in the public sector.

These AI contributions pose an exciting prospect which is coupled with a great responsibility for anticipating and mitigating potential harms. As with any novel technology, these risks and harms are not all immediately apparent.

In order to manage these impacts responsibly and to direct the development of AI systems toward optimal public benefit, considerations of AI ethics and governance must be a first priority. This will involve integrating considerations of the social and ethical implications of the design and use of AI systems into every stage of the delivery of AI/ML projects.
It will also involve a collaborative effort between the data scientists, product managers, data engineers, domain experts, and delivery managers to align the development of artificial intelligence technologies with ethical values and principles that safeguard and promote the wellbeing of the communities that these technologies affect.

> **KEY CONCEPT**
>
> **AI Ethics**
>
> The field of AI ethics tackles the social and moral implications of the production and use of AI technologies. It explores the values, principles, governance mechanisms, and deliberative processes needed to ensure the responsible and trustworthy design, development, deployment, and maintenance of AI systems.[4]

# Examples of the Use of AI Systems in the Public Sector

### Health and Social Care

Predicting the development of pandemics and epidemics to inform preventative interventions.[5]

Categorising children as 'at-risk' to inform decisions about the safety of the home environment.[6]

Predicting patients' risk in emergency rooms to triage patients or inform patient wait times.[7]

### Education

Automating assignment evaluations to save teachers' time and ensure consistency.[8] [9]

### Local Government

Predicting population trends (e.g. births) to inform development plans according to local needs.[10]

Predicting individuals' behaviour within services to inform interventions (e.g. encouraging individuals to save, to pay council taxes, to reduce antisocial behaviour).[11] [12] [13]

Identifying suitable sites for housing development.[14]

### Energy and Utilities

Predicting households' energy usage to inform personalised tariffs.[15] [16] [17]

Predicting maintenance needs and errors within energy generation and distribution systems to inform preventative action.[18]

### Transport

Predicting road maintenance needs to inform preventative action.[19]

Predicting traffic and controlling traffic signals to reduce congestion.[20]

Predicting long-term passenger needs across modes of travel to inform transport infrastructure plans.[21] [22]

**Environment and Agriculture**

Identifying sources contributing to air pollution to inform policy interventions.[23]

Predicting crops at risk of disease to inform appropriate treatment.[24]

**Defence and Security**

Predicting vulnerabilities within cybersecurity systems to inform preventative action.[25]

Predicting battlefield conditions and optimising military effectiveness by running simulations.[26]

**Criminal Justice**

Identifying individuals suitable for rehabilitation services to inform court decisions between rehabilitation and custody.[27]

Predicting individuals' risk of re-offence within the criminal justice system.[28]

Automating the analysis of digital evidence within court cases to streamline processes.[29]

**Immigration and Policing**

Predicting areas likely to have high criminal activity to inform police deployment.[30]

Categorising immigration applications (i.e. visa applications, residential status applications, citizenship applications) as low, mid, or high risk of being fraudulent to inform the level of human oversight over applications to streamline processes.[31] [32]

**Digital Markets and Communications**

Categorising online content to remove misinformation, disinformation, misleading advertisements, and scams.[33] [34]

**Across Domains**

Automating service provisions via conversational AI (e.g. licence approvals, customer service).[35]

Categorising users of digital public services to automate personalised content delivery (e.g. recommending relevant help articles to individuals using government platforms).[36] [37]

**Government**

Predicting optimal budget allocations and policies aimed to meet national strategy objectives.[38]

Key Concepts    Examples of AI in the Public Sector    13

# Technical Components of AI and Machine Learning (ML)

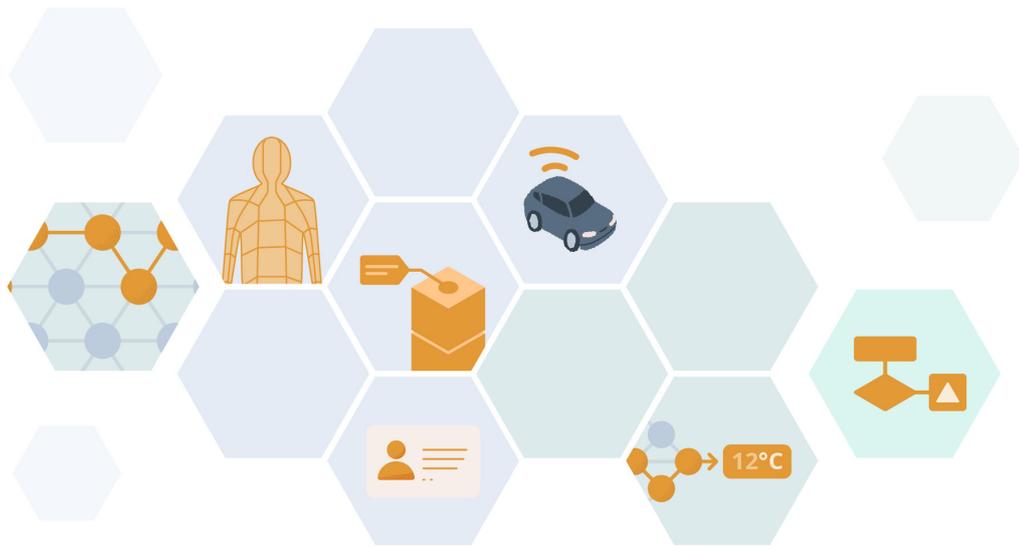

There are many ways to build AI/ML systems. Each involves the development of an algorithm that uses data to model some aspect of the world. This algorithm can then be applied to real-world data to make predictions about it. The definitions below are meant to demystify the two key components that make up AI systems: Data and Models .[39]

The technical concepts defined below have been drawn from many sources including publications such as Leslie, D., Burr, C., Aitken, M., Cowls, J., Katell, M., and Briggs, M. (2021). Artificial intelligence, human rights, democracy, and the rule of law: a primer. The Council of Europe. arXiv preprint arXiv:2104.04147 and ICO guidance (Information Commissioner's Office. (2021). Guide to the General Data Protection Regulation (GDPR). https://ico.org.uk/for-organisations/guide-to-data-protection/)

## Component 1
# Data

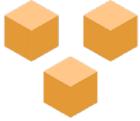
### Data

Data is the representation of information about the world recorded through observations (qualitative data) or measurements (quantitative data). In the context of AI, data is digitally recorded or transformed into a digital format.

**Examples of Data**

- Data relevant to environmental studies: average temperature readings from a weather station (numerical data), satellite images of land use (images), footage of wildfires (video), environmental impact statements (text), or traffic sounds (audio).

- Data relevant to healthcare: age or cholesterol level (numerical data), X-rays (images), endoscopy videos (video), clinical notes (text), or lung sounds recordings (audio).

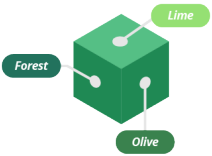
### Data Points

A data point is a discrete unit of information (or singular item of data).

**Examples of Data Points**

Data points contained within the feature of plant colour may include different colours such as red, orange, yellow, or pink, while those contained within the feature of place of origin may include different countries. Data points in the feature of height may take any continuous value in centimetres. A plant in the dataset that is red, originally from Malaysia and 28.5 cm tall has the data points 'red' for colour, 'Malaysia' for place of origin, and '28.5' for height.

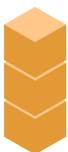
### Dataset

A collection of data that can range in type (e.g. numbers, words, images) and can either be utilised for a specific purpose or made available for general or multiple purposes.

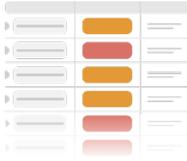
### Structured Dataset

A collection of data organised into tables with clearly defined categories.

**Examples of Structured Datasets**

- Spreadsheets of administrative and operational data.
- Official government statistics.[40]

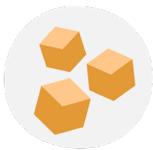
### Unstructured Dataset

A collection of general and varied data with no formatting or defined categories which must then be transformed into a computer-readable format.

**Examples of Unstructured Datasets**

- Collections of images and videos collected from CCTVs and satellites.
- Datasets scraped from social media websites.

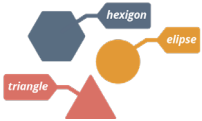
### Annotated Data

Structured or unstructured datasets where datapoints have been assigned labels which provide contextual information.

**Example of Annotated Data**

- In the case of a spam filter, a project team may use a dataset that contains emails labelled 'spam' or 'not spam'.
- If a project team was using an ML model to classify images as either being a cat or a dog, the dataset contains data labelled as 'cat', 'dog', 'neither'.

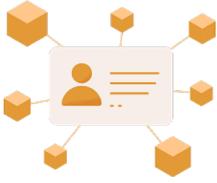
### Personal Data

Data that can be used to identify an individual. Some personal data need more protection because they are sensitive. These data are referred to as **special category data**.

**Examples of Personal Data**

- Official identity: name, identification numbers.
- Personal information: demographic information, work / education details, phone number, email addresses.
- Financial information: bank account details.
- Location: computer / phone IP address, GPS location.

**Examples of Personal Data That Are Also Special Category Data**

- Official identity: biometric data (fingerprints, facial patterns, voice) where used for identification purposes.
- Healthcare data: health records.
- Religious or political beliefs: party membership, voting history, and religious affiliation.

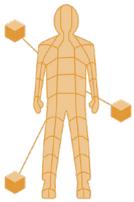
### Data Subject

A person whose personal data is collected, stored, processed, or used and who is identified or can be identified, directly or indirectly, by information such as a name or identity number, or by a combination of characteristics specific to that individual.

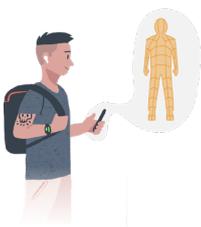
### Datafication

A process associated with the increasing use of technology, by which individuals, objects, and actions generate digital records (data).

**Example of Datafication**

The interactions of individuals with digital public-facing services generate data, such as demographic information, usage patterns, and user behaviour. Organisations that employ digital technologies to transform public facing services may collect, analyse, and use these data to inform future service delivery.

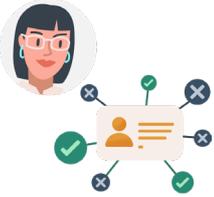
### Data Controller[41]

A natural or legal person, public authority, agency, or other body that, alone or jointly with others, exercises overall control over the purposes and means of the processing of personal data. This includes decisions over what data to process and why. In an AI project, data controllers play a crucial role in ensuring that an AI system is built and operated in a responsible and ethical manner. They must ensure data is collected, stored, processed, and used in accordance with relevant data protection regulations. Data processors are accountable for their own compliance and that of the processors.

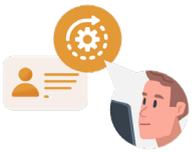
### Data Processor[42]

A natural or legal person, public authority, agency, or other body that does not have any purpose of their own for processing the data and only acts on the data controller's instructions.

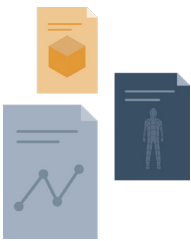
### Data Science

A field of study that includes elements from various disciplines including computer science, mathematics, statistics, and the social sciences, and is generally focused on extracting insights and patterns from datasets to answer or address a specific question or problem. Data science tools and technologies can be applied to other disciplines. These tools and technologies range from programming languages, statistical software, and data visualisation to big data platforms and machine learning libraries.

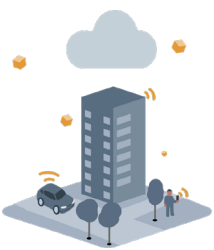
### Big Data

Datasets that are voluminous, contain a wide variety of data types, and are collected at great speed. They cannot be collected, managed, or analysed through traditional computational means, thereby requiring sophisticated digital tools and methods, as well as large amounts of storage. Big data can be used for revealing broad-scale patterns or trends.

**Examples of Big Data**

- Data generated from population interactions with digital public service platforms (e.g. HMRC or DWP).
- Datasets built by aggregating individual datasets generated within multiple public sector organisations.

<div style="text-align: center;">

**Component 2**
# Models

</div>

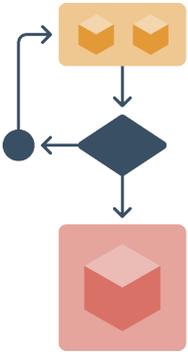

### Algorithm

A set of steps that are performed to complete a task or solve a problem. An algorithmic model is a formal representation (e.g. mathematical or logical) of the steps to be undertaken. Algorithms follow steps to map inputs to outputs. Humans could also follow an algorithmic process.

**Examples of Algorithms**

- When one bakes a cake by following a recipe, the recipe functions as an algorithm.

- When one solves an equation with a mathematical formula, the formula functions as an algorithm.

- A project team designing a navigation system uses a linear regression algorithm to predict estimated time of arrival (ETA). It first calculates the distance between its current location and the destination, and the average speed based on historical and current traffic patterns. It then determines the ETA by dividing the distance by the estimated speed. The ETA is updated as the vehicle moves towards the destination.

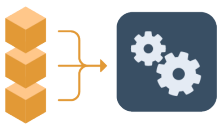

### Input Data

Input data refers to all data that are fed into a model to generate an output, from data used to train and test the model to the new data that results from the interaction of the AI system with the real-world.

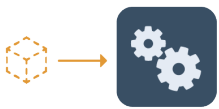

### Training Data

A subset of the dataset that is used to initially develop the model, by feeding the data into an algorithm.

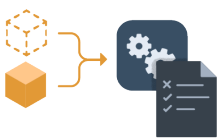

### Testing Data

Once a model has been trained, it is then tested on the remaining data (i.e. the subset of the dataset that was used for training the model). The data is split into training and testing data to ensure that the model can perform well for the original and newer data.

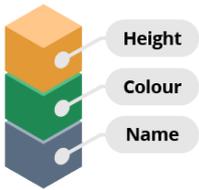

### Features

Features are characteristics within data that are organised into categories. Features are also referred to as variables and are the inputs that AI models use to generate outputs.

#### Examples of Features

- A structured dataset about plants may have features such as colour, plant height, name, and place of origin.
- Organised survey results.

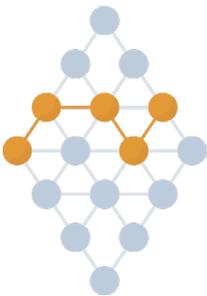

### Machine Learning (ML)

A popular approach to AI which uses training data to build algorithmic models which find patterns in that training dataset. Recent breakthroughs in AI have largely been in the realm of ML, therefore these workbooks primarily address the development of ethical ML-based AI systems.

When training is completed, ML models can ingest new or unseen data to predict outcomes for particular instances. "Learning" is a bit misleading, as the computer does not learn in the same way humans do. Instead, the computer uses mapping functions (i.e. mathematical formulas that map sets of inputs to sets of outputs). When new data are introduced during training ML models detect patterns contained in this new data and adjust their formulas accordingly to improve predictions.[43] Machine learning is used across many different sectors to model aspects of the world by applying algorithmic models to data and making inferences. With each iteration of their training, testing, and tuning, these models are honed to improve their ability to make correct predictions.

#### Example of Machine Learning

Video recommendation models map patterns in users' past watching history to predict what videos they are most likely to watch and to generate corresponding recommendations. The algorithm hones its parameters based on users' responses to these recommendations (i.e. watching or not watching a recommended video) and subsequently adjusts the formula for the generation of future recommendations.

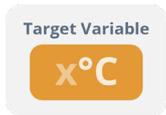

## Target Variable

When input data is labelled, it can be used to train algorithmic models, allowing them to "learn" to identify patterns that link inputs to outputs. The target variable is the intended output that an ML model tries to predict based on features in input data. Each of the features included in the model are used to predict the target variable.

### Example of a Target Variable

For an ML model built to predict what tomorrow's temperature may be, the target variable would be tomorrow's temperature. Other features in the dataset such as precipitation levels and humidity could be used by the model to predict this target variable.

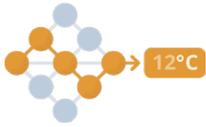

## Degrees of Automation

An AI system is fully automated if its output and any action taken as a result are implemented without any human involvement or oversight. In lower degrees of automation, a human can oversee the AI system to ensure it is producing the intended outcomes and/or use the outputs as part of a wider process in which they consider the output of the AI model, as well as other information available to them, and then acts based on this. This is often referred to as having a 'human-in-the-loop'. Degrees of automation can be seen as a spectrum rather than a binary concept and can vary depending on the specific context.

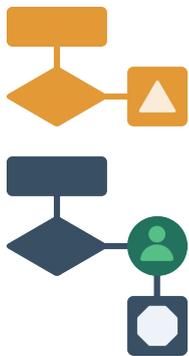

### Examples of Degrees of Automation

- **Full Automation -** An ML model used to prioritise patients for emergency care within an overcrowded hospital categorises a patient as ineligible for emergency care, automatically barring them access.

- **Human-On-The-loop -** An ML model categorises a patient as ineligible for emergency care. A nurse, who is aware of the patient's medical history and the patient's account of their current symptoms, considers the model's output and their own professional judgement. They decide the model's output is erroneous and override the model's decision.

- **Human-In-The-Loop -** An ML model categorises a patient as ineligible for emergency care. A nurse considers the model's output, the patient's medical history, the patient's account of their current symptoms, and their own professional judgement to determine the patient's eligibility.

- **Human-Over-The-Loop -** An ML model makes a preliminary determination about a patient's eligibility for emergency care. Before any decision is made, the ML model's output is automatically routed to a nurse, who reviews and approves or overrides the model's output before any action is taken.

# Example: Data Components in Practice

The Electoral Registration Officer (ERO) sends an annual electoral registration canvas to every residential address in the registration area. The canvas seeks to confirm the accuracy of electoral register details and identify new electors. But, there are several challenges to recieving responses. Firstly, only a small proportion of households return the form at the first stage. If households do not complete the form after a third reminder, a home visit may be necessary. Secondly, households that complete the form most often send it via a pre-paid envelope. The cost and time required for processing this data are higher than other available options (e.g. online form, SMS, and telephone).

To improve response rates, a County Council developed an AI system that analyses data on households in their borough. This system identifies individuals in the households who are **most likely to respond to the electoral registration canvas** and their **preferred way to send it**. The resulting insights help the local ERO to personalise messages in the electoral registration canvases they send out. This includes the wording used to describe the benefits of the registration and the letter design of the text indicating the options available to send the canvas.

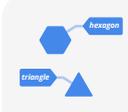

**Features**

Characteristics of the household members included in the dataset (e.g. household, location, citizenship status, and employment status) are used as the inputs into the model.

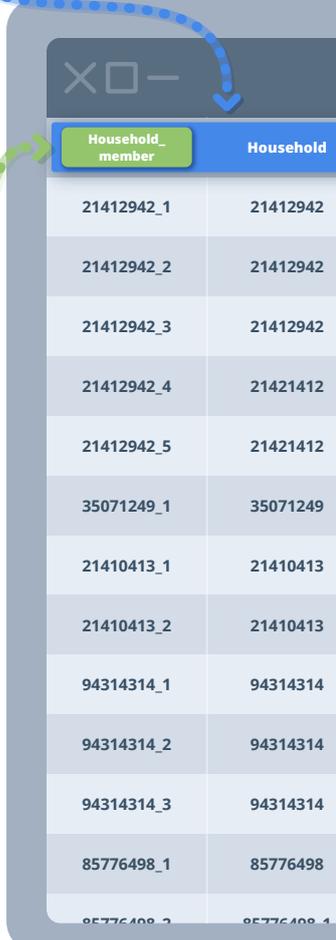

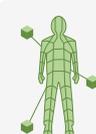

**Data Subject**

Household members' personal data is collected, stored, processed, and used in the AI system developed by the County Council. Household members can be identified, directly or indirectly, by a combination of information included in the database, such as postcode, location, age, and citizenship status.

### Personal Data

Postcode, location, age, citizenship status, annual income, employment status, and whether an individual voted in the last elections or not, can be used to identify the household members.

### Labelled Data

The dataset contains information on whether a household member has been registered to vote in the past and the preferred way to respond to the electoral registration canvas. This information serves as the clearly identified output (target variable or label).

### Structured Dataset

The spreadsheet represents a collection of data that is specific to the purpose of the AI system developed by the Local Council and organised into a table with clearly defined categories.

**AI system spreadsheet**

| Household_size | Postcode | Location | Age | Citizenship_status | Annual income | Employment_status | Voted_Last_Election | Prev_Electoral_Registration | Prev_communication |
|---|---|---|---|---|---|---|---|---|---|
| 5 | 32O L98 | Urban | 40 | Indefinite Leave to Remain | 50000 | Employed full-time | Yes | Yes | Pre-paid envelope |
| 5 | 32O L98 | Urban | 23 | British citizen | 0 | Student | No | No | N/A |
| 5 | 32O L98 | Urban | 35 | British citizen | 55000 | Self-employed | Yes | Yes | Pre-paid envelope |
| 5 | 32O L98 | Urban | 28 | Refugee | 28000 | Self-employed | Yes | Yes | Pre-paid envelope |
| 5 | 32O L98 | Urban | 32 | Indefinite Leave to Remain | 0 | Student | No | No | N/A |
| 1 | 87R P76 | Rural | 31 | British citizen | 15000 | Unemployed | Yes | Yes | Online form |
| 2 | 12L N13 | Rural | 59 | British Overseas Territories Citizens | 35000 | Self-employed | Yes | Yes | N/A |
| 2 | 12L N13 | Rural | 63 | Commonwealth Citizens | 12300 | Unemployed | No | No | N/A |
| 3 | 09E R12 | Rural | 18 | British citizen | 38000 | Employed full-time | No | No | N/A |
| 3 | 09E R12 | Rural | 15 | British citizen | 0 | Student | No | No | N/A |
| 3 | 09E R12 | Rural | 49 | Indefinite Leave to Remain | 76000 | Employed full-time | Yes | No | N/A |
| 4 | 98T W43 | Rural | 56 | British citizen | 23000 | Retired | No | No | SMS |
| 4 | 98T W43 | Rural | 46 | British Overseas | | Employed part-time | Yes | Yes | SMS |

### Data Point

Data points are individual units of information. When grouped together, a set of data points (i.e. a dataset) can help visualise or represent the possible values for a feature.

# A Closer Look at Machine Learning

Data-driven, machine learning ML models have emerged as one of the dominant forms of AI technology. These kinds of models may be constructed using a few different approaches; however, each of them involve finding patterns and drawing out inferences from data without explicit, formal instructions. The three main ML approaches are supervised, unsupervised, and reinforcement learning:[44]

## Supervised Learning

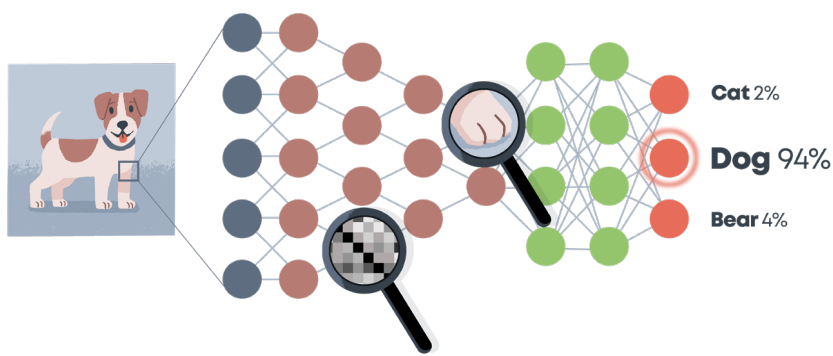

Supervised learning is the most widely used type of machine learning. Supervised ML models differ from unsupervised models because they are trained using labelled data. This means that the dataset contains 'labels' of the desired output or target variable that the model is trying to predict. Therefore, using this data, the model can find patterns between specific features in the dataset and the defined target variable (i.e. linking the inputs to the outputs). After the model is trained on this data, it can then be used to predict future outcomes by applying the model to new, unseen data that represents real-world scenarios.

For example, if a project team was using an ML model to classify images as either being a cat or a dog, the dataset used to train the model might contain various features that can be extracted from the image (e.g. lines, pixels, colours) and a labelled target variable (i.e. 'Cat', 'Dog', 'Neither'). During the training phase, the model identifies patterns between the various features of the image and what the image was classified as (i.e. the target variable). The model can then be applied during model testing to new images of cats and dogs and classify the new images based on the patterns discovered during the training phase.

Prior to the actual deployment of the AI model in real-world scenarios, the model is tuned and the team might revisit the model training process several times to ensure that correct predictions are being made on the new data and a desired standard of performance is being met. This process will be discussed in detail in the next section.

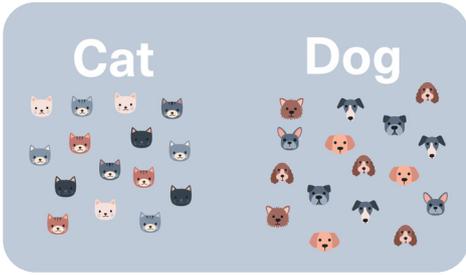
The tasks that Supervised ML models perform can include **classification** or **regression**. **Classification** models assign test data into specific categories (or classes) through prediction (such as grouping images by cat or dog, as illustrated).

**Regression** models determine the relationship between features and a target variable, such as predicting households' monthly energy usage through its relationship to a feature such as outside temperature.

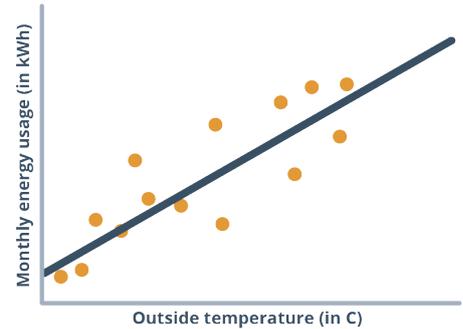

While linear regression and classification are the simplest forms of supervised learning, other supervised models such as support vector machines and random forests are also common applications.

# Unsupervised Learning

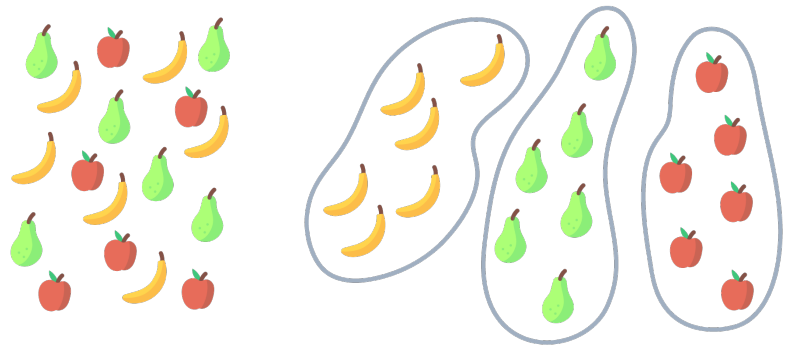

While supervised learning models map relationships between features in datasets that contain labels, unsupervised ML models identify patterns within unlabelled data by determining similarities and differences amongst the unlabelled data points. There are various applications of unsupervised learning, but one of the most common is clustering.

In clustering, a model receives unlabelled input data and determines similarities and differences among the input data points, resulting in clusters based on similar traits that are important factors in categorising the input data. In the image above, an ML model can be trained to separate images of fruits, animals, flowers, and trees into three separate clusters based on traits unique to each of the categories.

# Reinforcement Learning

Reinforcement learning models "learn" on the basis of their interactions with a virtual or real environment rather than existing data. Reinforcement learning "agents" search for an optimal way to complete a task by taking a series of steps that maximise the probability of achieving the given task. Depending on the success or failure of the steps they take, their actions are iteratively rewarded or penalised to maximise rewards. Reinforcement learning models improve with multiple iterations of trial and error to change their given "state". A "state" is the position of the agent at a specific point in time. It represents the current environment the agent is in and includes the collection of all relevant information that the agent needs to make decisions about what action to take next. Reinforcement learning models may be designed to develop long-term strategies to maximise their reward overall rather than looking only at their next step.

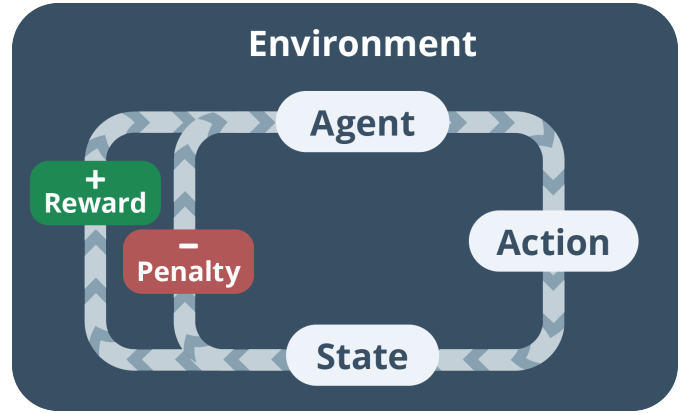

A common example of reinforcement learning can be found in the development of autonomous vehicles (self-driving cars). Reinforcement learning is used to improve the vehicle's performance in a simulated environment, testing for things such as response to traffic controls and acceleration. Through these interactions with the simulated environment, the reinforcement learning "agents" are penalised or rewarded based on task completion, thereby impacting the vehicle's future performance.

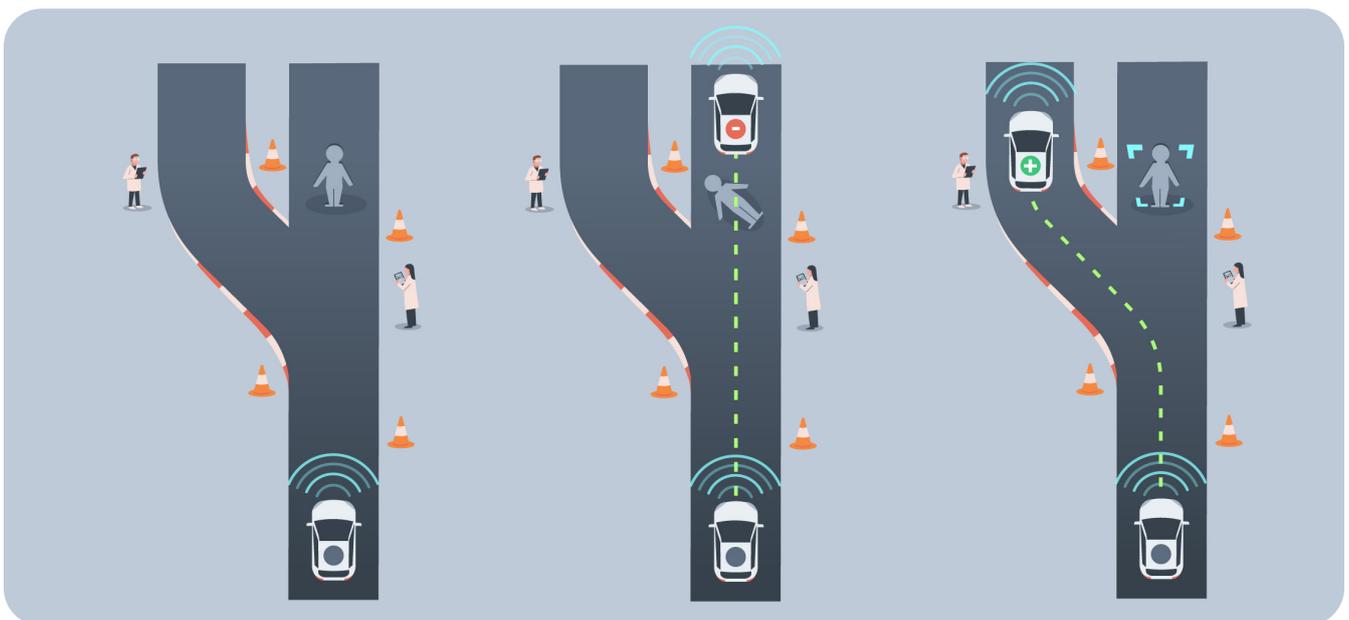

# Stages of the AI/ML Project Lifecycle

There are many ways of carving up the AI/ML project lifecycle*. However, it is important to work from a shared heuristic (an exploratory model) of the AI/ML project lifecycle that is optimally generalisable and practice-centred, while remaining algorithm neutral (i.e. that can apply to many different AI/ML techniques). It is also important that this approach prioritises the identification and mitigation of potential harms by taking into account the interwoven nature of its technical, social, and ethical aspects.

For the purposes of orienting readers of this workbook to the AI/ML workflow, we therefore use the model of the sociotechnical AI/ML project lifecycle developed in Burr and Leslie (2022),[45] which breaks down the AI/ML innovation workflow into stages of (Project) Design, (Model) Development, and (System) Deployment. Each of these overarching phases contains sub-phases that have been detailed in more depth. It is important to note that in addition to being a cyclical and iterative process, the sub-phases also naturally shade into one another. Often there is no clearly delineated boundary that differentiates certain project design activities (e.g. Data Extraction or Procurement and Data Analysis) from model design activities (e.g. Preprocessing & Feature Engineering, and Model Selection & Training), and Development activities from Deployment activities.

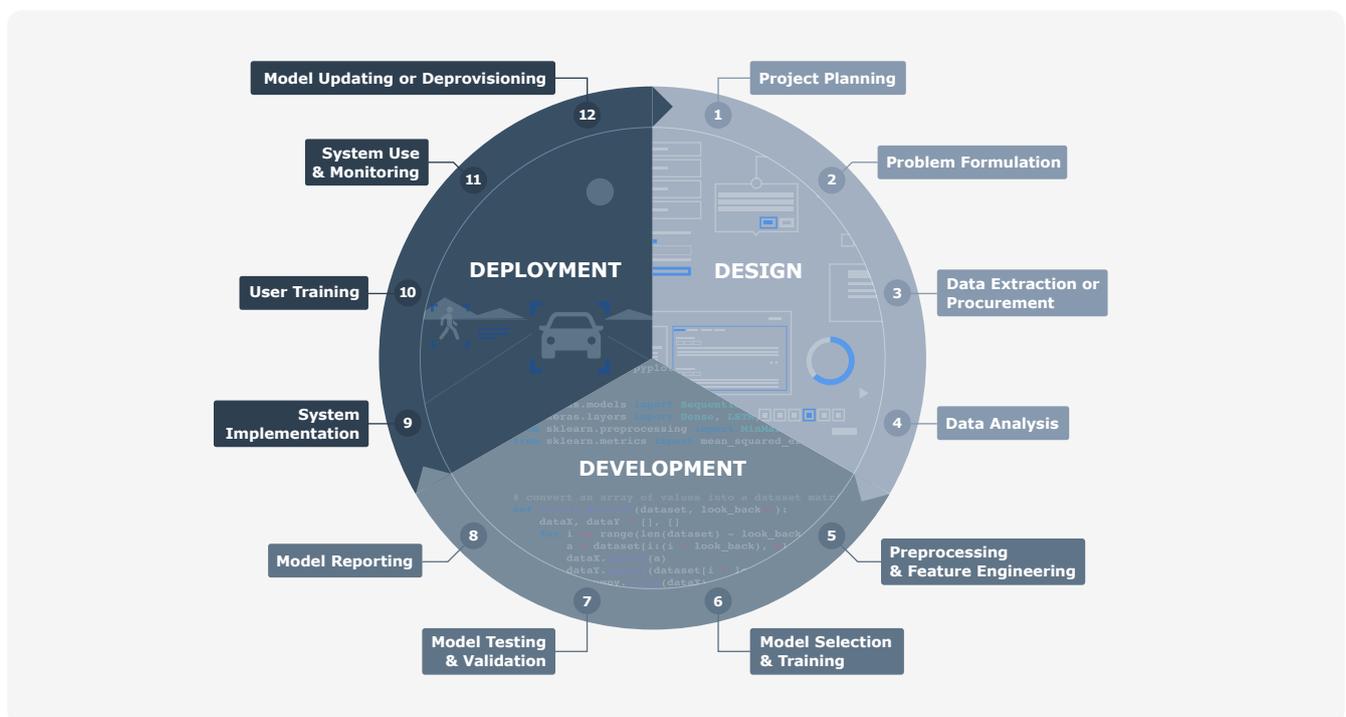

*Some examples of methods that aim to define the common tasks within a project lifecycle include Cross Industry Standard Process for Data Mining (CRISP-DM), the Knowledge Discovery in Databases (KDD) and Sample, Explore, Modify, Model, and Assess (SEMMA).[46] [47] [48]

**1** **Project Planning**

The project team must decide what the project's goals are at the outset. Tasks in this stage may include stakeholder analysis activities, initial project impact assessments, mapping of key stages within the project, or an assessment of resources and capabilities within the team or organisation.

**2** **Problem Formulation**

The project team needs to determine what problem their project is aiming to solve and how their desired output (target variable) should be defined and measured. If the team uses an indirect measure of the desired outcome, this is called a measurable proxy. Members of the team must also decide what input data are needed, given the problem to be solved, and for what purpose. The team should consider ethical and legal implications of the uses of data and provide a thorough account of intended and unintended consequences of use.

**3** **Data Extraction or Procurement**

This stage involves the processes by which data is gathered for the problem at hand. Data Extraction may involve web scraping processes or data recording through surveys or similar methodologies, whereas procurement may involve legal agreements to obtain already existing datasets.

**4** **Data Analysis**

At this stage, the project team can begin to inspect the data. Primarily, this will entail a high degree of exploratory data analysis (EDA). EDA involves understanding the makeup of the data through visualisation and summary statistics. Some questions at this stage may include: Are there missing data (incomplete data)? Are there outliers (unexpected data), unbalanced classes (imbalanced data), or correlations? As an iterative process, data analysis may inform and contribute to subsequent tasks of Data Extraction & Procurement.

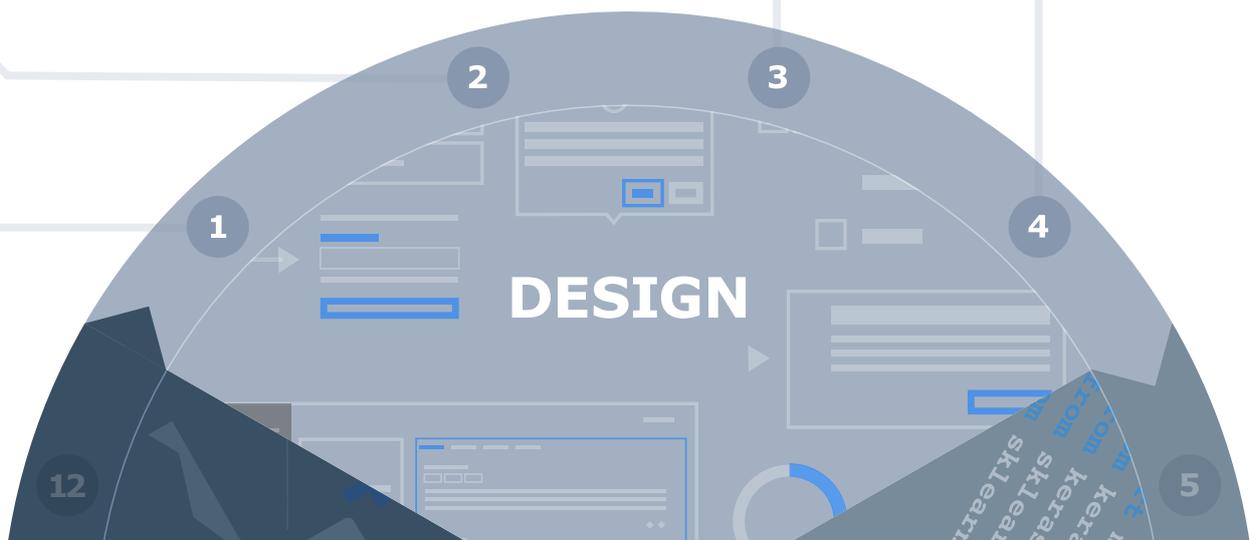

**DESIGN**

**5** **Preprocessing & Feature Engineering**

The Preprocessing stage is often the most time-consuming part of the development phase of the AI lifecycle. Preprocessing includes tasks such as data cleaning (reformatting or removing incomplete information), and data wrangling (transforming data into a format conducive for modelling amongst other processes that feed into the model training process).

**6** **Model Selection & Training**

Algorithms (mathematical formulas which form the basis of trained models) should be selected to serve the problem determined in the Design Phase. Model types vary in complexity; however, Model Selection considers other factors such as data types, quantity, and availability. Preprocessed data is split into a training set (around 60% of the data) testing set (20%) and validation set (20%), and the training dataset is used to hone the parameters of the selected model.

**7** **Model Testing & Validation**

After being trained, the model's performance (i.e. ability to deliver correct predictions) is tuned and tested with "testing" or "unseen" data, which has not been previously processed by the model. Validation sets are used to adjust higher-level aspects of the model (e.g. hyperparameters), and during this phase, changes can be made to the model's architecture to improve overall model performance.

**8** **Model Reporting**

After the team trains, tests, and validates the model, model evaluation (including a variety of performance measures) along with detailed information about the model workflow and re-visitation of the project's impact assessment should be produced to better support transparency and accountability.

## DEVENLOPMENT

## 9 System Implementation

At this stage, the project team can deploy the trained model in the real world. Effective implementation allows the model to be incorporated into a larger system. New data is processed by the implemented model to serve the intended purpose determined in the Design Phase.

## 10 User Training

Implementers of the system must be trained to understand the logic of the system, explain its decisions in plain language to decision subjects, and use independent and unbiased judgement to gauge the quality, reliability, and fairness of its outputs. Because of the iterative nature of the AI project lifecycle, user training may also inform and contribute to the effective implementation of the system.

## 11 System Use & Monitoring

During System Implementation, the team must monitor the model to ensure that it is still serving the desired purpose, being used responsibly and within the intended scope, and is responsive to emergent changes in real-world conditions that may lead to the model functioning less accurately or effectively. This entails conducting iterative performance re-evaluation and re-visitation of impact assessments throughout the span of deployment until model deprovisioning or retirement.

## 12 Model Updating or Deprovisioning

Over time, the model may lose efficacy, requiring the supervising team to revisit earlier stages of the development phase including model selection and training in order to make adjustments. If more significant changes are required, the system may need to be deprovisioned, thereby restarting at the design process with Project Planning.

## DEPLOYMENT

# Part Two: The Sociotechnical Aspect of the AI/ML Project Lifecycle

Understanding the technical stages of the AI/ML project lifecycle is an important first step in learning a practice-based approach to AI ethics and safety. The next step involves understanding the sociotechnical aspect of the AI/ML project workflow. This entails that we think about these project stages not simply as technical or scientific processes which are independent from the conditioning dynamics of the social environments in which they are embedded.[49]

Rather, from a sociotechnical perspective, technical processes are always, at the same time, social processes that are influenced by human values, interests, norms, and institutions. They are processes that are often shaped by potentially biased human choices, potentially fallible human judgments, and potentially unjust social structures.[50]

> **KEY CONCEPT**
>
> **The Sociotechnical Aspect of the AI/ML Project Lifecycle**
>
> The AI/ML project lifecycle is affected by the intertwined relation between technology and social structures (e.g. the people in the AI project and the environment in which the project team operates). Both elements interact and influence each other. To treat the design, development, and deployment of data-driven technologies as both social and technical constructs is to adopt a sociotechnical approach to the AI/ML project lifecycle.

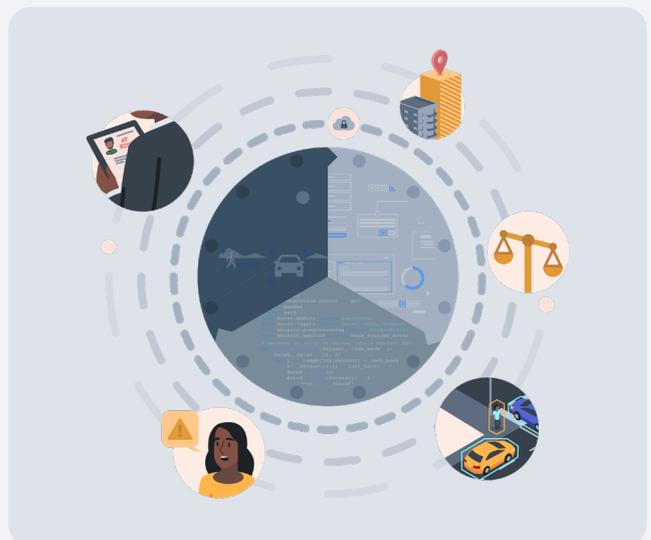

Acknowledging this human shaping of AI technology is important, because it prompts us to recognise that, from the earliest stages of the sociotechnical project lifecycle, human choices, interests, and values are integrated into AI Project Design, Model Development, and System Deployment:

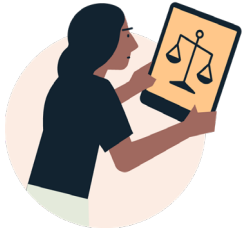

### PROJECT DESIGN

At the Project Planning stage, for instance, human judgements need to be made about whether building an algorithmic model is the right approach given available resources and data, existing technologies and processes already in place, the complexity of the use-contexts involved, and the nature of the social problem that needs to be solved. These path-determining choices wield agenda-setting power in AI innovation ecosystems—a power that is all-too-often hoarded by those who control resources and that is thus exercised in ways that are at cross-purposes with goals of democratic governance, public consent, and social license.[51] [52]

### KEY CONCEPT

**Agenda-Setting Power**

Agenda-setting power refers to the power of an individual or organisational actor over another actor to the extent that the former sets the agenda that the latter must follow. The actor with agenda-setting power delimits the range of possibilities for the behaviour of the actor it has power over to those that they consider acceptable, tolerable, or desired. These possibilities frequently fall into the sphere of their influence and interest.

### MODEL DEVELOPMENT

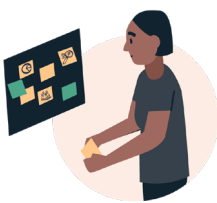

At the Problem Formulation stage, human evaluations and interests also shape the determination of what target variables should be implemented within the system. This means that the very acts of devising the statistical problem and of translating project goals into measurable proxies can introduce structural biases which may ultimately lead to discriminatory harm.[53]

Likewise, at the Preprocessing & Feature Engineering stage, human decisions about how to group or separate input features (e.g. how to carve up categories of gender or ethnic groups) or about which input features to exclude altogether (e.g. leaving out deprivation indicators in a predictive model for clinical diagnostics) can have significant influences on the fairness and equity of AI/ML systems.

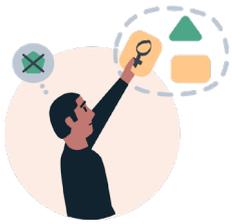

**SYSTEM DEPLOYMENT**

At the System Deployment stage, insufficiently trained implementers may introduce their own cognitive biases into the use of AI technologies. For example, an implementer of an automated decision support system may become hampered in their critical judgment and situational awareness as a result of their faith in the perceived objectivity, neutrality, or superiority of the system. Likewise, an implementer, who works in a safety critical domain where liability considerations are prioritised, may fear the consequences of departing from the recommendation of an AI system even where their judgement of the situation leads them to a different course of action. Both of these cases may lead to over-compliance or errors of commission where implementers begin to unreflectively defer to the perceived authority or infallibility of the system and thereby become unable to detect problems emerging from its use for reason of a failure to hold the results against available information and common sense judgement.

Being able to discern these sociotechnical pinch-points across the project lifecycle is crucial for AI ethics, because it enables us to recognise that the activities steering AI innovation processes are ethically implicated. Such activities are charged with a responsibility for critical self-reflection about the role that human purposes, values, biases, and interests play at every juncture of the AI project workflow. Likewise, from the sociotechnical perspective, those involved in such activities bear a responsibility to anticipate and consider the real-life impacts that these AI innovation processes yield.

This awareness that all AI innovation processes possess sociotechnical aspects and ethical stakes directs AI project teams to incorporate end-to-end habits of responsible research and innovation (RRI) into all of their research and innovation activities. From a RRI viewpoint, creating a culture of responsible AI innovation involves putting into practice habits of continuous and contextually-responsive critical reflection and deliberation from the first moments of project design straight through to system retirement.[54] We will now distill these RRI habits into what we call the CARE and Act Framework.

> **KEY CONCEPT**
>
> **Responsible Research and Innovation**[55]
>
> Responsible Research and Innovation (RRI) is a framework for reflecting on, anticipating, and deliberating about the ethical, social, and legal questions that arise in the research and development of scientific and technological tools, practices, and systems. This framework provides tools and methods for identifying and evaluating the potential harms and opportunities of data-driven technologies and addressing challenges in a reasonable and socially acceptable manner. In this sense, Responsible Research and Innovation entails the adoption of a forward-looking approach to responsibility.

# The CARE and Act Framework

As a starting point for thinking about how to incorporate habits of RRI into the project lifecycle, we introduce the CARE and Act Framework. Considering the CARE and Act maxims at each stage of the AI/ML project lifecycle can help project teams in establishing and sustaining habits of RRI.[56]

> **KEY CONCEPT**
>
> **CARE and Act Framework**
>
> A tool for sense checking and reflecting on the values, purposes, and interests that steer AI/ML projects, as well as projects' real-world implications.

### C  Consider Context

Think about the conditions and circumstances surrounding your research and innovation project. Focus on the norms and interests that motivate your team and your community of practice: How are these influencing or steering your project? Take into account the specific domain in which it is situated and reflect on the expectations of affected stakeholders that derive therefrom.

### A  Anticipate Impacts

Describe and analyse the impacts, intended or not, that might arise from your project. This does not seek to predict but rather to support an exploration of possible risks and implications that may otherwise remain uncovered and little discussed.

### R  Reflect on Purposes, Positionality, and Power

Reflect on the goals of and motivations for the project; Scrutinise the way that potential perspectival limitations could exercise influence on the equity and integrity of the project; Reflect on the power imbalances that could exist between the project team and the communities impacted by your project: How, if at all, could these power imbalances result in the injustice or harm?

### E  Engage Inclusively

Open up such visions, impacts, and questioning to broader deliberation, dialogue, engagement, and debate in an inclusive way. Embrace peer review at all levels and welcome different views.

### Act  Act Transparently and Responsibly

Use these processes to influence the direction and trajectory of the research and innovation process itself. Produce research that is both scientifically and ethically justifiable (EPSRC, 2013, expanded) and provide transparent documentation, following best governance, self-assessment, and reporting practices.

# AI Ethics and Governance in Practice

While the CARE and Act Framework is intended to orient and enable general habits of RRI, the AI Ethics and Governance in Practice Programme that will be presented in this workbook series will aim to bring such habits to life. It will do this by focusing on the specific governance processes and actions needed to put ethical values and practical principles into practice across the AI project lifecycle. This will involve introducing and describing a multi-tiered governance framework that enables project teams both to reflectively integrate such values and principles into their innovation practices and to have clear mechanisms for demonstrating and documenting this.

The recently adopted standard, ISO 37000, defines governance as 'the system by which the whole organisation is directed, controlled, and held accountable to achieve its core purpose in the long run'. Establishing a diligent and well-conceived governance framework that covers the entire design, development, and deployment process will provide the foundation for your team to steward the design and implementation of AI projects towards responsible delivery. While many ethical frameworks exist, we present here the Process Based Governance (PBG) Framework outlined by Understanding AI Ethics and Safety.[57] This governance architecture adopts ethically sound practices at every point in the project lifecycle. It is composed of three building blocks, with each level building on from the last:

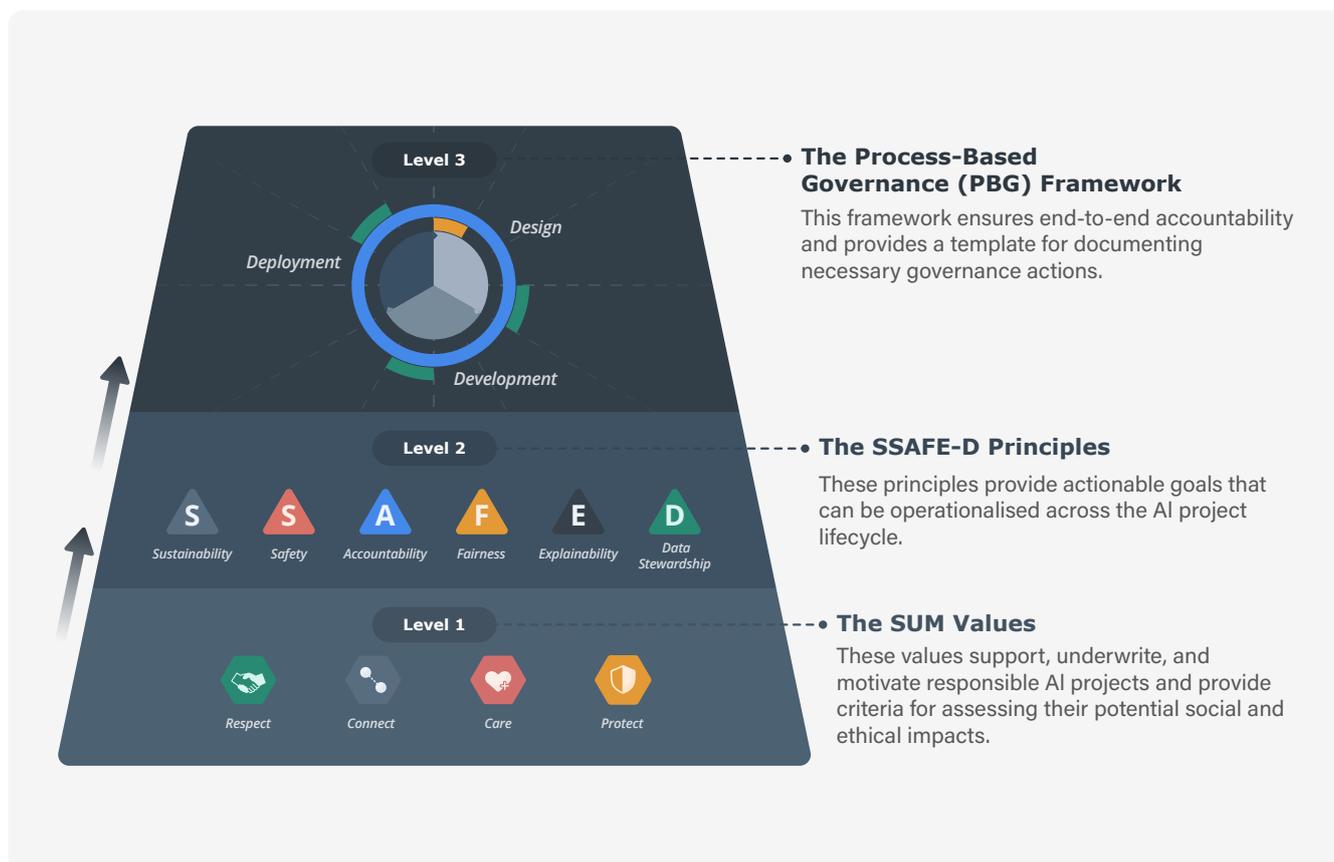

**The Process-Based Governance (PBG) Framework**
This framework ensures end-to-end accountability and provides a template for documenting necessary governance actions.

**The SSAFE-D Principles**
These principles provide actionable goals that can be operationalised across the AI project lifecycle.

**The SUM Values**
These values support, underwrite, and motivate responsible AI projects and provide criteria for assessing their potential social and ethical impacts.

| Level 1 | Level 2 | Level 3 |
|---|---|---|
| The **SUM Values** are a set of ethical values intended to provide general normative guideposts and moral motivations for thinking through the social and ethical aspects of AI projects, and to establish well-defined criteria to evaluate their potential impacts as well as their ethical permissibility. | At the second, and more concrete level, the **SSAFE-D Principles** are a set of actionable goals which provide tools to make sure that AI projects are bias-mitigating, non-discriminatory, and fair, and that they safeguard public trust in their capacity to deliver safe, sustainable, accountable, transparent, and reliable AI innovation. | At the third and most concrete level, the **Process-Based Governance (PBG) Framework** provides a mechanism for integrating the SSAFE-D Principles within AI design, development, and deployment processes as well as a template for documenting corresponding governance actions. |

### Level 1
# The SUM Values

The PBG Framework is built from the cultural ground up, with its foundation being the SUM Values. These values **support**, **underwrite**, and **motivate** a responsible AI projects. They provide an accessible framework of criteria for considering, assessing, and deliberating on the **potential impacts** and **ethical permissibility** of a prospective AI project and are meant to be utilised as **guiding values throughout the innovation lifecycle**:

**KEY CONCEPT**

**Values**
Basic and fundamental beliefs that guide and motivate attitudes or actions.

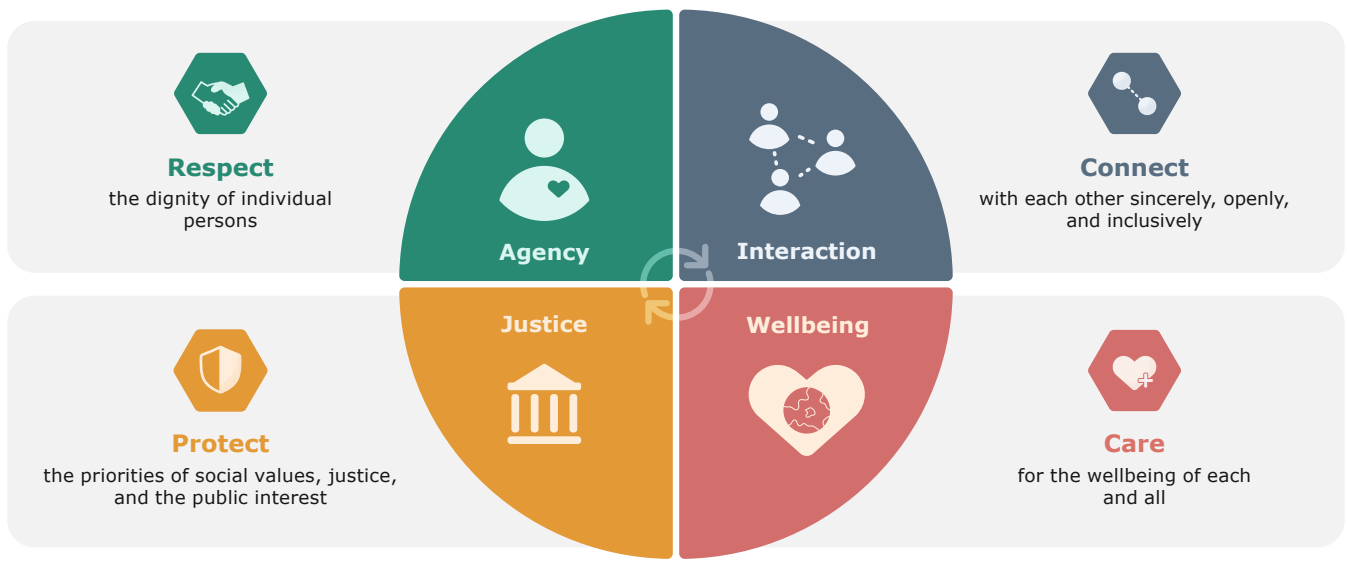

**Respect** the dignity of individual persons

**Connect** with each other sincerely, openly, and inclusively

**Protect** the priorities of social values, justice, and the public interest

**Care** for the wellbeing of each and all

Agency | Interaction
Justice | Wellbeing

The values of respect, connect, care, and protect respond to the set of real-world risk posed by the use of the AI/ML technologies.

| Risks that Emerge From the Use of AI/ML Technologies | Related Ethical Implications and Concerns | SUM Values |
|---|---|---|
| Loss of autonomy | **Agency** — Human agency, dignity, and individual flourishing | **Respect** the dignity of individual persons |
| Loss of interpersonal connection and empathy | **Interaction** — Solidarity, communication, and integrity of social interaction | **Connect** with each other sincerely, openly, and inclusively |
| Poor quality or hazardous outcomes | **Wellbeing** — Individual, communal, and biospheric wellbeing | **Care** for the wellbeing of each and all |
| Bias, injustice, inequality, and discrimination | **Justice** — Justice, equity, and the common good | **Protect** the priorities of social values, justice, and the public interest |

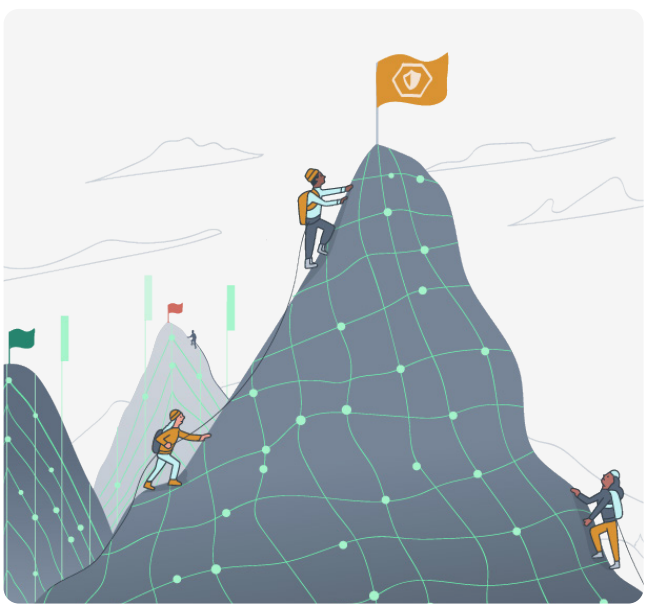

The SUM Values are meant to be utilised as guiding values throughout the AI lifecycle. Adopting common values from the outset of projects enables respectful, sincere, and open dialogue about ethical challenges by providing members of project teams with a shared vocabulary and criteria for assessing the ethical permissibility of AI projects. In keeping with the SUM Values, project teams can be united in a collaborative spirit to develop AI technologies for the public good.

| Level 2 |
|---|

# The SSAFE-D Principles

While the SUM Values are intended to provide general guideposts for thinking through the social and ethical aspects of AI project delivery, they do not detail how to ensure that these values shape the design, development, and deployment of AI systems. Building on the SUM Values are therefore the SSAFE-D Principles: practical goals that can be established and achieved for AI projects to make sure that projects are ethically justifiable.

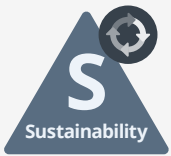 Achieving this goal requires assuring AI projects being developed with continuous sensitivity to real-world impacts.

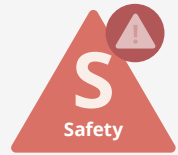 Achieving this goal requires an AI system to be technically accurate, reliable, secure, and robust.

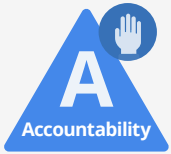 Achieving this goal requires assuring the project's end-to-end answerability and auditability.

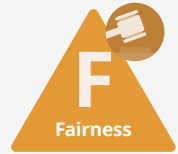 Achieving this goal requires assuring a minimum threshold of discriminatory non-harm and bias mitigation.

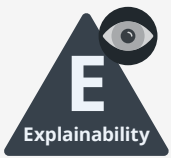 Achieving this goal requires the ability to explain and justify AI project processes and AI-supported outcomes.

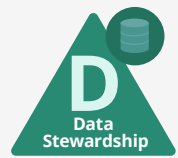 Achieving this goal requires data quality, integrity, protection, and privacy to be assured.

Putting principles in motion and ensuring they are achieved in practice requires the implementation of **governance actions**.

Across these workbooks, each of the SSAFE-D principles is accompanied by corresponding governance actions. We present a list of these governance actions on the following page.

> **KEY CONCEPT**
>
> **Governance Action**
> Processes aimed at operationalising principles and achieving normative goals.

# List of Principles and Governance Actions

## Sustainability

**SEP (Stakeholder Engagement Process)**
Process facilitating the uptake of proportionate stakeholder engagement and input throughout the AI lifecycle. The SEP enables a contextually informed understanding of the social environment and human factors that may be impacted by, or may impact, individual AI projects.[58]

**SIA (Stakeholder Impact Assessment)**
Process facilitating the iterative evaluation of the social impact and sustainability of individual AI projects, as well as the corroboration of these potential impacts in dialogue with stakeholders, when appropriate.

## Accountability

**PBG Log**
Live document outlining governance actions, relevant team members and roles involved in each action, timeframes for follow-up actions, and logging protocols, for individual AI projects.[59]

## Explainability

**EAM (Explainability Assurance Management)**
Iterative process aimed to facilitate the implementation and evaluation of transparency and explainability assurance activities across the project lifecycle and assist in providing clarification of AI system outputs to a range of impacted stakeholders.

## Safety

**SSA & RM (Safety Self-Assessment and Risk Management)**
Process facilitating the evaluation of how AI projects align with safety objectives through the iterative identification and documentation of potential safety risks across the lifecycle and assurance actions implemented to address these.

## Fairness

**BSA, BRM & FPS (Bias Self-Assessment, Bias Risk Management & Fairness Position Statement)**
Process facilitating the evaluation of how AI projects align with the principle of fairness through the iterative identification and documentation of risks of bias across the lifecycle and assurance actions implemented to address these. The FPS is a document establishing the metric-based fairness criteria for individual AI projects, providing an explanation in plain and nontechnical language.

## Data Stewardship

**Data Factsheet**
Live document facilitating the uptake of best practices for responsible data management and stewardship across the AI project workflow. The document consists of a comprehensive record of the data lineage and iterative assessments of data integrity, quality, protection, and privacy.

# The Process-Based Governance (PBG) Framework

The purpose of the PBG Framework is to ensure that the entirety of the SSAFE-D Principles are successfully operationalised and documented across the AI project lifecycle. It is a template that provides a landscape view of where in the AI project workflow governance actions are to take place in order to integrate each of the SSAFE-D principles within AI project activities.

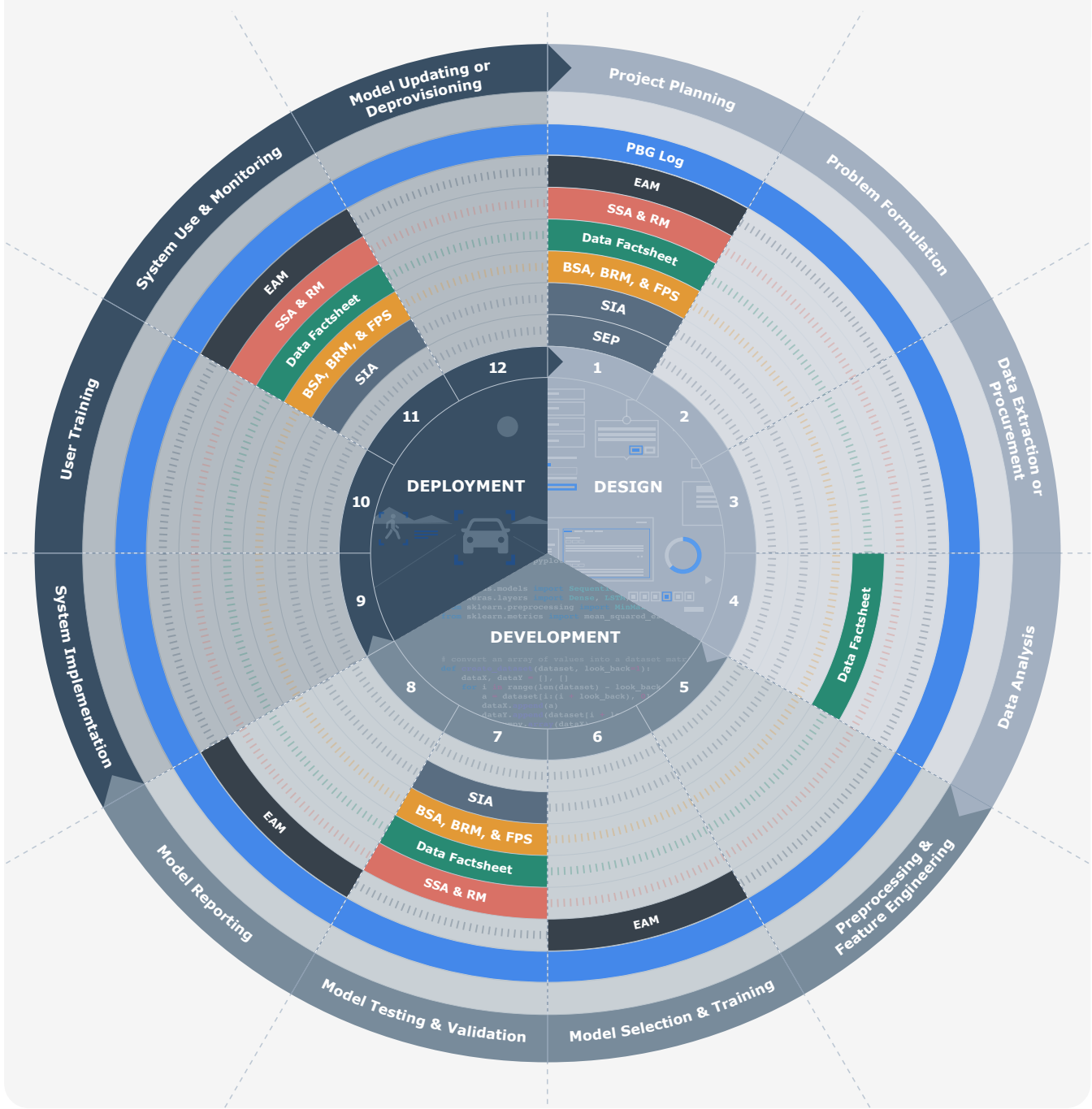

The governance actions integrated into individual AI projects will, in practice, be tailored to the needs of each specific project as teams establish governance actions that are contextually appropriate and proportional to the potential risks and hazards presented by their project. In all instances, however, the effective implementation of the PBG Framework will result in the production of a process log or PBG Log. This is used to organise and document the governance actions established to integrate the SSAFE-D Principles into the design, development, and deployment of a given AI system. The PBG Log allows project teams to demonstrate and document the successful operationalisation of each principle. It functions as a detailed register of governance actions and stewards the end-to-end transparency and accountability of AI projects, providing a documentary touchpoint for ensuring that AI systems are produced and used ethically, safely, and responsibly.

**A PBG Log Provides Documentation Of:**

- Established governance actions across the project lifecycle.
- Relevant team members and roles involved in each governance action.
- Explicit timeframes for follow-up actions, re-assessments, and continual monitoring.
- Clear and well-defined protocols for logging activity and instituting mechanisms for end-to-end audibility.

**Implementing a PBG Log Requires Two Steps:**

- Establishing appropriate governance controls and actions for a project.
- Logging project activities based on the established governance controls.

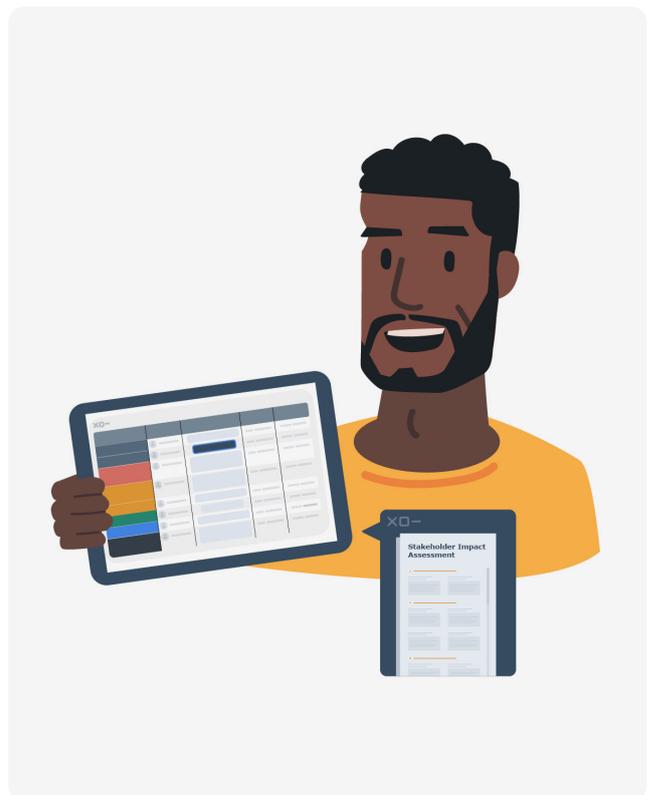

Effective implementation of the PBG Framework will result in the production of a process log. The process log functions as a detailed register of governance actions that stewards the end-to-end transparency and accountability of AI projects, providing a documentary touchpoint for ensuring that AI systems are produced and used ethically, safely, and responsibly.

**WB 8** More details about the PBG Log can be found in **Workbook 8: AI Accountability in Practice.**

AI Ethics and Governance in
Practice: An Introduction

# Activities

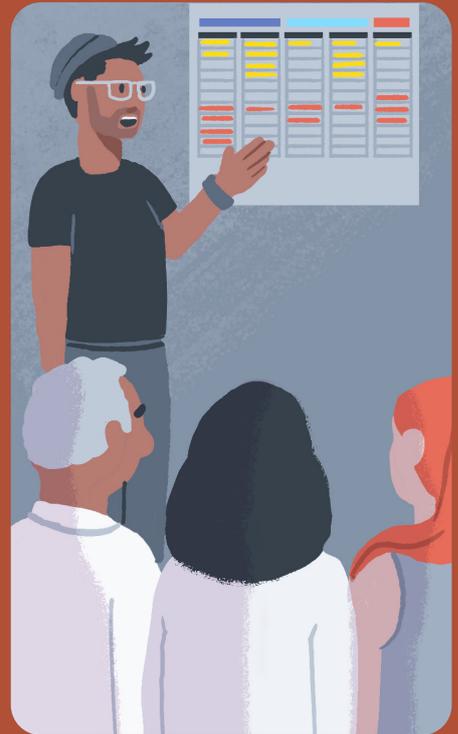

# Activities Overview

In the previous sections of this workbook, we have presented an introduction to the core concepts of the AI and Ethics and Governance in Practice Programme. In this section we provide concrete tools for applying these concepts in practice. Activities will help demystify AI and ML by exploring the key technical components that make up AI systems, different types of ML, the AI lifecycle model, and an introduction to Responsible Research and Innovation.

We offer a collaborative workshop format for team learning and discussion about the concepts and activities presented in the workbook. To run this workshop with your team, you will need to access the resources provided in the link below. This includes a Miro board with case studies and activities to work through.

Workshop resources for **AI Ethics and Governance in Practice (An Introduction):** turing.ac.uk/aieg-1-activities

**A Note on Activity Case Studies**

Case studies within the Activities sections of the AI Ethics and Governance in Practice workbook series offer only basic information to guide reflective and deliberative activities. If activity participants find that they do not have sufficient information to address an issue that arises during deliberation, they should try to come up with something reasonable that fits the context of their case study.

> **Note for Facilitators**
>
> In this section, you will find the participant and facilitator instructions required for delivering activities corresponding to this workbook. Where appropriate, we have included Considerations to help you navigate some of the more challenging activities.
>
> Activities presented in this workbook can be combined to put together a capacity-building workshop or serve as stand-alone resources. Each activity corresponds to a section within the Key Concepts in this workbook. Some activities have pre-requisites, which are detailed on the following page.

### Collective Image

Illustrate the team's collective image of AI.

**Corresponding Sections**

→ Introduction to Artificial Intelligence (page 10)

→ A Closer Look at Machine Learning (page 24)

→ Technical Components of AI/ML (page 14)

### Exploring Public Sector AI

Consider public sector AI and participants' potential status as data subjects.

**Corresponding Sections**

→ Examples of the Use of AI Systems in the Public Sector (page 12)

### Mapping the AI/ML Project Lifecycle

Develop an understanding of the different steps taken in designing, developing, and deploying AI systems.

**Corresponding Sections**

→ Stages of the AI/ML Project Lifecycle (page 27)

### Exploring the CARE and Act Framework

Understand how the CARE and Act Framework can serve as a starting point for establishing habits of critical reflection and responsible innovation across the AI/ML project lifecycle.

**Corresponding Sections**

→ The Sociotechnical Aspect of the AI/ML Project Lifecycle (page 31)

### Revisiting Collective Image

Revisit the team's collective image of AI.

**Pre-Requisites**

↗ Activity: Collective Image (page 45)

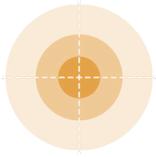

35 mins | Participant Instructions

# Collective Image

**Objective**

The purpose of this activity is to develop an understanding of the team's collective perception of AI.

**Team Instructions**

1. Your team will divide into groups. Each group is to discuss what they think AI is and come up with a visual representation. Consider the questions below and deliberate in your groups box in the **Visual Representation** section:

   - Can your group think of some examples of AI technologies?
   - Are AI systems present in team members' lives or work?
   - Who is involved in developing AI systems?
   - Who is affected by the design, development, and use of AI systems?

2. When indicated by your facilitator, reconvene as a team and share your visual representations.

3. Have a team deliberation about the common characteristics of your representations.

4. Individually place sticky notes on the **Confidence Levels** and **Responses Diagram** sections. Each team member will share how they feel about AI and how confident they are with their understanding of what AI is.

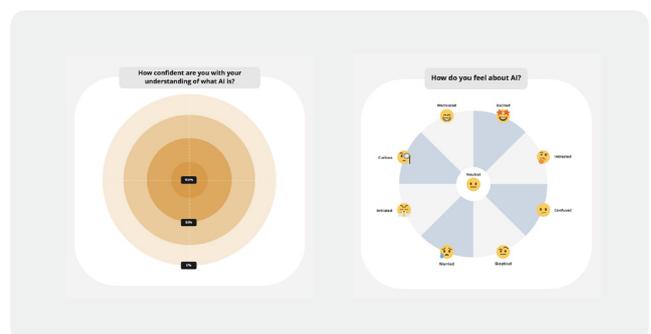

**Confidence Levels** and **Responses Diagram**

*This activity was inspired by the work of Lewis, T., Gangadharan, S. P., Saba, M., Petty, T. (2018). Digital defense playbook: Community power tools for reclaiming data. Detroit: Our Data Bodies.*[60]

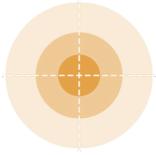

# Collective Image

1. Give the group a moment to read over the activity instructions.

2. Divide the team into groups, giving the team 15 minutes to discuss what they think AI is and come up with a visual representation. Ask each group to consider the question 'What is "AI"?'. They can also consider the following questions:

   - Can your group think of some examples of AI technologies?
   - Are AI systems present in team members' lives or work?
   - Who is involved in developing AI systems?
   - Who is affected by the design, development, and use of AI systems?

3. Their visual representations are meant to illustrate their perceptions of AI. Emphasise that there are no wrong answers. Ask them to feel free to make use of examples, images, and texts to illustrate their representation.

   - **If delivering physically:** encourage the team to make use of the workshop materials provided.
   - **If delivering digitally:** encourage the team to make use of the illustration tools within your digital board and online image searches.

4. When 15 minutes have passed, ask the team to reconvene and share their visual representations.

5. Lead a team deliberation about the common characteristics of their illustrations.

   - **Co-facilitator:** Write common characteristics in notes under **Key Characteristics** within the **Collective Image** section of the board.

6. Ask the team to individually place sticky notes on the **Confidence Levels** and **Responses Diagrams** on the board to share how they feel about AI and how confident they are in their understanding of what AI is.

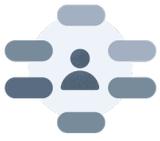

45 mins | Participant Instructions

# Exploring Public Sector AI

**Objective**

The purpose of this activity is for team members to consider public sector AI and participants' potential status as data subjects.

**Team Instructions**

1. Your team will be divided into groups, each with a volunteer note-taker that will report back to the team.

2. Each group is to consider an assigned example of AI in the public sector, deliberating what data may be processed by this model. Consider the questions:

    - What information could be useful for answering the question this model is trying to address?
    - What personal data might have been processed by this model?
    - How might this data be created and collected?

3. Reconvene as a whole team. Each note-taker will report back to the team, considering the questions:

    - What was the group's example?
    - What data might have been used in this model and for what purpose?
    - What personal data might have been processed by this model?

4. Take a brief look at other examples of public sector AI uses, having a whole team discussion about how your data might interact with these systems.

---

**KEY CONCEPT**

**Personal Data**

Data that can be used to identify a living individual. Examples include official identity documents, personal information, healthcare data, financial information, location, religious or political beliefs. Find out more in the Technical Components of AI and ML section on page 14 of this workbook.[61]

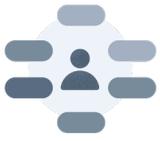

45 mins | Facilitator Instructions

# Exploring Public Sector AI

1. Give the team a moment to read over the instructions for this activity. Answer any of their questions.

2. Divide the team into groups, each with an assigned Public Sector AI Use Case.
   - Ask for a volunteer note-taker for each group. Volunteers will report back on group discussions to the greater team.

3. Give each group 20 minutes to consider their assigned AI system, using the questions in the instructions to deliberate possible data processed by their models.

4. When the time has passed, ask the groups to reconvene with the whole team.

5. Give each group 2 minutes to share about their discussions.

6. After each group has shared, lead a team discussion about public sector AI. Ask the team to consider these and other Examples of AI in the Public Sector and consider:
   - Who are the data subjects in these models?
   - What components of their personal data might be processed by some of these models?

7. Write the team's comments in sticky notes, placing them in the **Notes** section of the board.

**Public Sector AI Use Cases**

| Use Cases | Data Source / Types |
|---|---|
| **Health and Social Care:** ML model used to monitor compliance with social distancing guidelines within a pandemic context. | Anonymised phone location data, live camera feeds, and urban footfall data (collected by street sensors). |
| **Education:** ML model used to identify students as at risk of not completing their education to inform interventions. | Age, gender, education history, and housing status. |
| **Local Government:** ML model used to categorise household types for local councils to inform targeted promotion of public services. | Census data, property data, residents' family history, and education levels. |

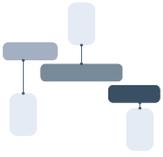

45 mins | Participant Instructions

# Mapping the AI/ML Project Lifecycle

**Objective**

The purpose of this activity is to help you develop an understanding of the different steps taken in designing, developing, and deploying AI systems.

**Team Instructions**

1. Your team will be divided into groups. Each group will be assigned a phase in the AI/ML project lifecycle. Take a moment to review the steps within your assigned stage.

2. Next, turn to the **Mapping the AI/ML Project Lifecycle** section. Each step of the AI/ML project lifecycle is outlined in order. Each step is connected to activities that together recount how an AI system was designed, developed, and deployed. **Activity descriptions, however, are not linked to the correct steps.**

3. With your group, re-organise the activities in your assigned phase, connecting them to their corresponding steps.

4. Next, assign an activity to each group member, ensuring all activities in your phase are assigned.

5. Reconvene with the rest of the workshop team. Team members are to read out their assigned steps and activities in order.

6. Individually take a moment to review the narrative of this AI system, considering how it reflects each step of the AI/ML project lifecycle and asking any questions.

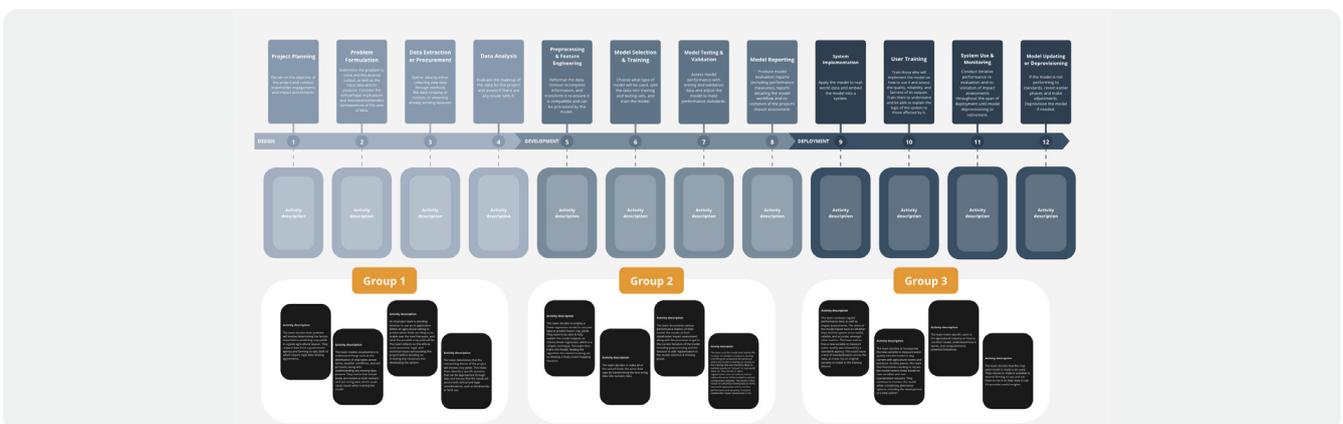

**Mapping the AI/ML Project Lifecycle**

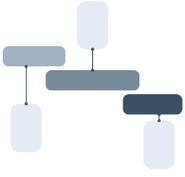

45 mins | Facilitator Instructions

# Mapping the AI/ML Project Lifecycle

1. Give the team time to read the objective and team instructions for this activity. Ask the team if they have any questions.

2. Divide the team into groups, each assigned a phase in the AI/ML project lifecycle.

3. Give the groups 5 minutes to review the steps within their assigned phase.

4. Next, give the groups 15 minutes to re-organise the sequence of activities in the diagram.

5. When the time is up, ask the team to stop working on the diagram.

6. Ask the group to reconvene.

7. Lead the groups into reading out their assigned steps and activities in order (i.e. Step 1. Project Planning, 'A project team is deciding on…').

8. If any of their answers are incorrect or missing, read over the correct steps and activities with the team, allowing them to ask any questions and answering with the Considerations section of this activity.

    - **Co-facilitator:** Adjust the diagram if this occurs.

9. Give the team a moment to look over the diagram, asking for questions a second time and using the Considerations section to respond.

# DESIGN

| Step Number, Name & Definition | Activity Description |
|---|---|
| **1** **Project Planning** Decide on the objective of the project and conduct stakeholder engagements and impact assessments. | An AI project team is deciding whether to use an AI application within an agricultural setting to predict which fields are likely to be arable over the next five years, and what the possible crop yield will be. The team reflects on the ethical, socio-economic, legal, and technical issues surrounding this project before deciding on investing any resources into developing the system. |
| **2** **Problem Formulation** Determine the problem to solve and the desired output, as well as the input data and its purpose. Consider the ethical/legal implications and intended/unintended consequences of the uses of data. | The team determines that the overarching theme of the project will involve crop yields. This helps them identify a specific question that can be approached through data and ensure that the result will accord with ethical and legal considerations, such as biodiversity or land use. |
| **3** **Data Extraction or Procurement** Gather data by either collecting new data through methods like data scraping or surveys, or obtaining already existing datasets. | The team decides their problem will involve determining the factors important in predicting crop yields in a given agricultural season. They request data from a government agency and farming co-ops, both of which require legal data sharing agreements. |
| **4** **Data Analysis** Evaluate the makeup of the data for the project and assess if there are any issues with it. | The team creates visualisations to understand things such as the distribution of crop types across farms, weather conditions, and soil pH levels, along with understanding any missing data present. They notice that soil pH levels are treated as both numeric and text string data, which could cause issues when training the model. |

# DEVELOPMENT

| Step Number, Name & Definition | Activity Description |
|---|---|
| **5** **Preprocessing & Feature Engineering** Reformat the data, remove incomplete information, and transform it to ensure it is compatible for processing by the model. | The team decides to make all of the soil pH levels the same data type by transforming the text string data into numeric data. |
| **6** **Model Selection & Training** Choose what type of model will be used, split the data into training and testing sets, and train the model. | The team decides to employ a linear regression model to use past data to predict future crop yields. They want to be able to fully explain the model outputs, so choose linear regression, which is a simpler technique. The team then trains the model, feeding the algorithm the cleaned training set to develop a finely tuned mapping function. |
| **7** **Model Testing & Validation** Assess model performance with testing and validation data and adjust the model to meet performance standards. | The team runs the model and realises the number of variables included is causing overfitting (an unwanted ML behaviour where the model is hewing too closely to the training data and therefore likely to perform poorly on "unseen" or real-world data). So, they decide to add a regularisation term (a method used to reduce the error of the model) to remove unimportant variables. The model is then tested on unfamiliar testing data to mimic real world application and to confirm performance and accuracy. A second stakeholder impact assessment is conducted. |
| **8** **Model Reporting** Produce model evaluation reports (including performance measures), reports detailing the model workflow and re-visitation of the project's impact assessment. | The team documents various performance metrics of their model, the results of their stakeholder impact assessment, along with the processes to get to the current iteration of the model including preprocessing and the decision to add regularisation in the model selection & training phase. |

# DEPLOYMENT

| Step number, name & definition | Activity description |
|---|---|
| **9 System Implementation** Apply the model to real-world data and embed the model into a system. | The team decides that the crop yield model is ready to be used. They choose to make it available to several farming co-ops and ask them to run it on their data to see if it provides useful insights. |
| **10 User Training** Train those who will implement the model on how to use it and assess the quality, reliability, and fairness of its outputs. Train them to understand and be able to explain the logic of the system to those affected by it. | The team trains specific users in the agricultural industry on how to use their model, understand how it works, and comprehend its potential limitations. |
| **11 System Use & Monitoring** Conduct iterative performance re-evaluation and re-visitation of impact assessments throughout the span of deployment until model deprovisioning or retirement. | The team conducts regular performance tests as well as impact assessments. The users of the model report back on whether they find the system to be useful, reliable, and accurate, amongst other metrics. The team notices that a new variable to measure water quality was released by a standards agency. This could cause a lack of standardisation across the data, as it was not an original variable included in the training dataset. |
| **12 Model Updating or Deprovisioning** If the model is not performing to standards, revisit earlier phases and make adjustments. Deprovision the model if needed. | The team decides to incorporate the new variable to measure water quality into the model to stay current with agriculture norms and practices. As time passes, the team find themselves needing to retrain the model several times based on new variables and non-standardised datasets. They continue to monitor the model while considering alternative options, including the development of a new system. |

# Exploring the CARE and Act Framework

**Objective**

The purpose of this activity is to understand how the CARE and Act Framework can serve as a starting point for establishing habits of critical reflection and responsible innovation across the AI/ML project lifecycle.

**Case Study**

Your organisation's website hosts a popular digital public forum. The forum has a high volume of online participation. To ensure the mitigation of potential online harms, your team is launching an AI system that automatically detects abusive language. This type of tool can be a crucial protective measure against hate crimes, cyberbullying, and discriminatory language. However, due to the rapid evolution of offensive colloquialisms, the tool may not detect new forms of abusive language that were not included in its original training dataset. If this system is not updated regularly, it may not detect instances of online harm caused by abusive language.

> Please note: This case study offers basic information aimed to guide group reflection.

**Team Instructions**

1. Your team will be divided into groups, each with a volunteer note-taker that will report back to the team. All groups will be assigned the same case study, but each group will focus on a different principle contained in the CARE and Act Framework.

2. Read the case study as a group, discussing the possible ethical challenges presented by the case study.

3. Next, read the description of your group's assigned case study discussing how this principle could be important in addressing the challenges your group has identified by answering the question provided. Use sticky notes to write down your answers.

4. Next, in your groups, discuss the two columns outlined beneath your principle, using sticky notes to write down your thoughts.

5. Reconvene as a group and reflect, considering:

   - What would the implications of not putting each of the CARE and Act principles into practice be?

   - What would the benefits of putting each of the CARE and Act principles into action be?

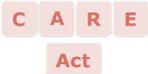

| 45 mins | Facilitator Instructions |

# Exploring the CARE and Act Framework

1. Give the team a moment to read over the instructions and case study. Answer any of their questions.

2. Divide the teams into groups, each assigned a principle in the CARE and Act framework.

3. Give each group 20 minutes to discuss possible ethical challenges presented by the case study and how their assigned principle could be important in addressing this challenge.

4. When the allocated time has passed, ask the groups to reconvene.

5. Give each group 2 minutes to share their scenarios and responses.

6. After each group has shared, lead a discussion considering the questions in the Participant Instructions.

   - **Co-facilitator:** Take notes on the team discussion, writing these in the **Discussion Notes** section.

Facilitator Instructions    Exploring the CARE and Act Framework    55

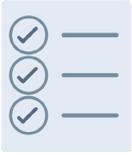

⏱ 20 mins  |  Participant Instructions

# Revisiting Collective Image

**Objective**

Revisit team's collective perception of AI.

**Team instructions**

1. Individually take a moment to consider the role of ethics in the AI/ML project lifecycle and look over the results of the **Collective Image activity**, considering the extent to which you believe AI ethics may help address negative emotions within the **Responses Diagram**.

2. Fill out the **Revisited Collective Image** section.

3. Have a group discussion about how, if at all, your collective image of AI has changed through this workshop.

**Revisited Collective Image**

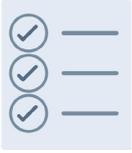

20 mins | Facilitator Instructions

# Revisiting Collective Image

1. Give the group a moment to read over the activity instructions and look at the **Collective Image** section of the board.

2. Ask the team to consider the extent to which they believe AI ethics may help address challenging sentiments within the **Collective Image** section of the board. Give them five minutes to individually review the Collective Image diagram, filling out the diagrams within the **Revisited Collective Image** section and **Workshop Feedback** sections by using notes to write out and share their viewpoints.

3. When 5 minutes have passed, ask the team to reconvene.

4. Lead a group discussion, considering the following questions:
   - Has the team's level of confidence in understanding AI changed? If so, how?
   - Has the team's responses to AI changed? If so, how?

# Endnotes

1. Kumar, Y., Koul, A., Singla, R., & Ijaz, M. F. (2022). Artificial intelligence in disease diagnosis: a systematic literature review, synthesizing framework and future research agenda. *Journal of Ambient Intelligence and Humanized Computing,* 1-28. https://doi.org/10.1007/s12652-021-03612-z

2. Daut, M. A. M., Hassan, M. Y., Abdullah, H., Rahman, H. A., Abdullah, M. P., & Hussin, F. (2017). Building electrical energy consumption forecasting analysis using conventional and artificial intelligence methods: A review. *Renewable and Sustainable Energy Reviews, 70,* 1108-1118. https://doi.org/10.1016/j.rser.2016.12.015

3. Jia, Y., Weiss, R. J., Biadsy, F., Macherey, W., Johnson, M., Chen, Z., & Wu, Y. (2019). Direct speech-to-speech translation with a sequence-to-sequence model. *arXiv preprint arXiv:1904.06037.*

4. Leslie, D. (2019). Understanding artificial intelligence ethics and safety. *arXiv preprint arXiv:1906.05684.*

5. Zeng, D., Cao, Z., & Neill, D. B. (2021). *Artificial intelligence–enabled public health surveillance—from local detection to global epidemic monitoring and control.* In Artificial intelligence in medicine (pp. 437-453). Academic Press. https://doi.org/10.1016/B978-0-12-821259-2.00022-3

6. Leslie, D., Holmes, L., Hitrova, C., & Ott, E. (2020). *Ethics of machine learning in children's social care.* Zenodo. http://doi.org/10.5281/zenodo.3676569

7. Tahayori, B., Chini-Foroush, N., & Akhlaghi, H. (2021). Advanced natural language processing technique to predict patient disposition based on emergency triage notes. *Emergency Medicine Australasia, 33*(3), 480-484. https://doi.org/10.1111/1742-6723.13656

8. Baker, T., Smith, L., & Anissa, N. (2019). *Educ-AI-tion Rebooted? Exploring the future of artificial intelligence in schools and colleges.* 56. https://media.nesta.org.uk/documents/Future_of_AI_and_education_v5_WEB.pdf

9. Dencik, L., Hintz, A., Redden, J. and Warne, H. (2018) *Data Scores as Governance: Investigating uses of citizen scoring in public services.* Research Report. Cardiff University. https://datajusticelab.org/wp-content/uploads/2018/12/data-scores-as-governance-project-report2.pdf

10. Symons, T. (2016). *Datavores of Local Government.* Nesta. https://www.nesta.org.uk/report/datavores-of-local-government/

11. Symons, T. (2016). *Datavores of Local Government.* Nesta. https://www.nesta.org.uk/report/datavores-of-local-government/

12. Griffiths, H. (2016). *IoT Adoption Among Cities in the UK (27).* IotUK. https://iotuk.org.uk/wp-content/uploads/2016/08/IoT_Adoption_Security_Report.pdf

13. Local Government Association. (n.d.) *Behavioural insights: resources and best practice.* https://www.local.gov.uk/our-support/behavioural-insights/behavioural-insights-resources-and-best-practice

14  Urban Intelligence. (2021, July 13). *Birmingham using AI to find land for homes.* https://urbanintelligence.co.uk/news/birmingham-using-ai-to-find-land-for-homes/

15  Chou, J. S., & Tran, D. S. (2018). *Forecasting energy consumption time series using machine learning techniques based on usage patterns of residential householders.* Energy, 165, 709-726. https://doi.org/10.1016/j.energy.2018.09.144

16  Wang, Z., & Srinivasan, R. S. (2017). A review of artificial intelligence based building energy use prediction: Contrasting the capabilities of single and ensemble prediction models. *Renewable and Sustainable Energy Reviews, 75,* 796-808.

17  Esnaola-Gonzalez, I., Jelić, M., Pujić, D., Diez, F. J., & Tomašević, N. (2021). *An AI-powered system for residential demand response.* Electronics, 10(6), 693. https://doi.org/10.3390/electronics10060693

18  Rhodes, A. (2020). *Digitalisation of energy: An Energy Futures Lab briefing paper.* Energy Futures Lab. https://doi.org/10.25561/78885

19  Symons, T. (2016). *Datavores of Local Government.* Nesta. https://www.nesta.org.uk/report/datavores-of-local-government/

20  Soomro, S., Miraz, M. H., Prasanth, A., & Abdullah, M. (2018). *Artificial intelligence enabled IoT: traffic congestion reduction in smart cities.* IET Conference Proceedings, IET Digital Library. https://digital-library.theiet.org/content/conferences/10.1049/cp.2018.1381

21  Ubaldi, B., Le Fevre, E. M., Petrucci, E., Marchionni, P., Biancalana, C., Hiltunen, N., Intravaia, D. M., & Yang, C. (2019). *State of the art in the use of emerging technologies in the public sector* (31). OECD Publishing. https://doi.org/10.1787/932780bc-en

22  Berryhill, J., Heang, K. K., Clogher, R., & McBride, K. (2019). *Hello, World: Artificial intelligence and its use in the public sector.* OECD Working Papers on Public Governance (36), OECD Publishing. https://doi.org/10.1787/726fd39d-en.

23  Ubaldi, B., Le Fevre, E. M., Petrucci, E., Marchionni, P., Biancalana, C., Hiltunen, N., Intravaia, D. M., & Yang, C. (2019). *State of the art in the use of emerging technologies in the public sector* (31). OECD Publishing. https://doi.org/10.1787/932780bc-en

24  Ubaldi, B., Le Fevre, E. M., Petrucci, E., Marchionni, P., Biancalana, C., Hiltunen, N., Intravaia, D. M., & Yang, C. (2019). *State of the art in the use of emerging technologies in the public sector* (31). OECD Publishing. https://doi.org/10.1787/932780bc-en

25  Burke, A. (2020). *Robust artificial intelligence for active cyber defence* (31). Centre for Data Ethics and Innovation. https://www.gov.uk/government/publications/cdei-ai-barometer

26  Ministry of Defence (2021). *Defence in a competitive age* (CP411). https://assets.publishing.service.gov.uk/government/uploads/system/uploads/attachment_data/file/974661/CP411_-Defence_Command_Plan.pdf

27 Oswald, M., Grace, J., Urwin, S., & Barnes, G. C. (2018). Algorithmic risk assessment policing models: lessons from the Durham HART model and 'Experimental'proportionality. *Information & communications technology law, 27*(2), 223-250. http://shura.shu.ac.uk/17462/

28 Rigano, C. (2019). Using artificial intelligence to address criminal justice needs. *National Institute of Justice Journal, 280*(1-10), 17. https://nij.ojp.gov/topics/articles/using-artificial-intelligence-address-criminal-justice-needs

29 Privacy International. (2021). *Digital stop and search: how the UK police can secretly download everything from your mobile phone.* https://privacyinternational.org/sites/default/files/2018-03/Digital%20Stop%20and%20Search%20Report.pdf

30 Burke, A. (2020). *Robust artificial intelligence for active cyber defence* (31). Centre for Data Ethics and Innovation. https://www.gov.uk/government/publications/cdei-ai-barometer

31 Home Office. (2018). *Biometrics Strategy Better public services Maintaining public trust.* https://assets.publishing.service.gov.uk/government/uploads/system/uploads/attachment_data/file/720850/Home_Office_Biometrics_Strategy_-_2018-06-28.pdf

32 Home Office. (2019). *Fourth report on statistics being collected under the exit checks programme.* https://assets.publishing.service.gov.uk/government/uploads/system/uploads/attachment_data/file/826381/fourth-report-on-statistics-being-collected-under-the-exit-checks.pdf

33 Cambridge Consultants. (2019). *Use of AI in Online Content Moderation: 2019 report produced on behalf of Ofcom.* https://www.ofcom.org.uk/__data/assets/pdf_file/0028/157249/cambridge-consultants-ai-content-moderation.pdf

34 Department for Digital, Culture, Media, and Sport & Home Office. (2020). *Online Harms White Paper: Full government response to the consultation.* https://www.gov.uk/government/consultations/online-harms-white-paper/outcome/online-harms-white-paper-full-government-response

35 Central Digital and Data Office. (2020). *Using chatbots and webchat tools: How to use chatbots and webchat tools to improve your users' experience of your service.* https://www.gov.uk/guidance/using-chatbots-and-webchat-tools

36 Burke, A. (2020). *Robust artificial intelligence for active cyber defence.* Alan Turing Institute: Defence and Security Programme. https://www.turing.ac.uk/sites/default/files/2020-08/public_ai_acd_techreport_final.pdf

37 Ubaldi, B., Le Fevre, E. M., Petrucci, E., Marchionni, P., Biancalana, C., Hiltunen, N., Intravaia, D. M., & Yang, C. (2019). *State of the art in the use of emerging technologies in the public sector* (31). OECD Publishing. https://doi.org/10.1787/932780bc-en

38 Bennett, S., & Cutler, N. (2019, October 28). Lab Long Read: Policy Consultations - Part 2: A role for data science?. *Policy Lab and Department for Transport (DfT).* https://openpolicy.blog.gov.uk/2019/10/28/lab-long-read-policy-consultations-part-2-role-of-data-science/

39  This section was inspired by the work of Payne, B. H. (2019). An ethics of artificial intelligence curriculum for middle school students. *MIT Media Lab Personal Robots Group.* https://thecenter.mit.edu/wp-content/uploads/2020/07/MIT-AI-Ethics-Education-Curriculum.pdf

40  For instance, refer to datasets provided by the Office for National Statistics. https://www.ons.gov.uk/

41  Information Commissioner's Office. (2021). *Guide to the General Data Protection Regulation (GDPR)*. https://ico.org.uk/for-organisations/guide-to-data-protection/

42  Information Commissioner's Office. (2021). *Guide to the General Data Protection Regulation (GDPR)*. https://ico.org.uk/for-organisations/guide-to-data-protection/

43  Leslie, D., Burr, C., Aitken, M., Cowls, J., Katell, M., and Briggs, M. (2021). Artificial intelligence, human rights, democracy, and the rule of law: a primer. The Council of Europe. *arXiv preprint arXiv:2104.04147*

44  This section draws from the work of Leslie, D., Burr, C., Aitken, M., Cowls, J., Katell, M., and Briggs, M. (2021). Artificial intelligence, human rights, democracy, and the rule of law: a primer. The Council of Europe. *arXiv preprint arXiv:2104.04147*

45  Burr, C., & Leslie, D. (2022). Ethical assurance: a practical approach to the responsible design, development, and deployment of data-driven technologies. *AI and Ethics,* 1-26. https://doi.org/10.1007/s43681-022-00178-0

46  Sweenor, D., Hillion, S., Rope, D., Kannabiran, D., Hill, T., & O'Connell, M. (2020). *ML Ops: Operationalizing Data Science.* O'Reilly Media, Incorporated. https://www.oreilly.com/library/view/ml-ops-operationalizing/9781492074663/

47  Wirth, R., & Hipp, J. (2000). CRISP-DM: Towards a standard process model for data mining. In *Proceedings of the 4th international conference on the practical applications of knowledge discovery and data mining.* 1, 29-3. http://www.cs.unibo.it/~danilo.montesi/CBD/Beatriz/10.1.1.198.5133.pdf

48  Shafique, U., & Qaiser, H. (2014). A comparative study of data mining process models (KDD, CRISP-DM and SEMMA). *International Journal of Innovation and Scientific Research, 12*(1), 217-222. https://www.issr-journals.org/playoffs/ijisr/0012/001/IJISR-14-281-04.pdf

49  Burton, S., Habli, I., Lawton, T., McDermid, J., Morgan, P., & Porter, Z. (2020). *Mind the gaps: Assuring the safety of autonomous systems from an engineering, ethical, and legal perspective.* Artificial Intelligence, 279, 103201. https://doi.org/10.1016/j.artint.2019.103201

50  Burr, C., & Leslie, D. (2022). Ethical assurance: a practical approach to the responsible design, development, and deployment of data-driven technologies. *AI and Ethics,* 1-26. https://doi.org/10.1007/s43681-022-00178-0

51  Bachrach, P., & Baratz, M. S. (1962). Two faces of power1. *American political science review, 56*(4), 947-952. https://doi.org/10.2307/1952796

52  Leslie, D., Katell, M., Aitken, M., Singh, J., Briggs, M., Powell, R., Rincon, C., Perini, A. M., Jayadeva, S., & Mazumder, A. (2022). *Advancing Data Justice Research and Practice: An Integrated Literature Review.* arXiv preprint https://arxiv.org/pdf/2204.03090
AI Ethics and Governance in Practice
An Introduction

53 Leslie, D., Katell, M., Aitken, M., Singh, J., Briggs, M., Powell, R., Rincon, C., Perini, A. M., Jayadeva, S., & Burr, C. (2022). *Data Justice in Practice: A Guide for Developers*. arXiv preprint https://arxiv.org/pdf/2205.01037

54 Burr, C., Fischer, C., and Rincon, C. (2023) *Responsible Research and Innovation (Turing Commons Skills Track).* The Alan Turing Institute. https://doi.org/10.5281/zenodo.7755693

55 Burr, C., Fischer, C., & Rincon, C. (2023) *Responsible Research and Innovation (Turing Commons Skills Track).* The Alan Turing Institute. https://doi.org/10.5281/zenodo.7755693

56 Leslie, D. (2020). Tackling COVID-19 Through Responsible AI Innovation: Five Steps in the Right Direction. *Harvard Data Science Review,* (Special Issue 1). https://doi.org/10.1162/99608f92.4bb9d7a7

57 Leslie, D. (2019). *Understanding artificial intelligence ethics and safety: A guide for the responsible design and implementation of AI systems in the public sector.* The Alan Turing Institute. https://doi.org/10.5281/zenodo.3240529

58 Leslie, D., Katell, M., Aitken, M., Singh, J., Briggs, M., Powell, R., Rincon, C., Perini, A. M., & Jayadeva, S. (2022). *Data Justice in Practice: A Guide for Policymakers.* SSRN. https://dx.doi.org/10.2139/ssrn.4080050

59 Leslie, D. (2019). *Understanding artificial intelligence ethics and safety: A guide for the responsible design and implementation of AI systems in the public sector.* The Alan Turing Institute. https://doi.org/10.5281/zenodo.3240529

60 Lewis, T., Gangadharan, S. P., Saba, M., Petty, T. (2018). *Digital defense playbook: Community power tools for reclaiming data. Detroit: Our Data Bodies.* https://www.odbproject.org/wp-content/uploads/2019/03/ODB_DDP_HighRes_Single.pdf

61 Leslie, D., Burr, C., Aitken, M., Cowls, J., Katell, M., and Briggs, M. (2021). Artificial intelligence, human rights, democracy, and the rule of law: a primer. The Council of Europe. *arXiv preprint arXiv:2104.04147s*

62 Burr, C., & Leslie, D. (2022). Ethical assurance: a practical approach to the responsible design, development, and deployment of data-driven technologies. *AI and Ethics,* 1-26. https://doi.org/10.1007/s43681-022-00178-0

# Bibliography and Further Readings

## Technical Background

Alom, M. Z., Taha, T. M., Yakopcic, C., Westberg, S., Sidike, P., Nasrin, M. S., Hasan, M., Van Essen, B.C., Awwal, A.A.S., & Asari, V.K. (2019). A State-of-the-Art Survey on Deep Learning Theory and Architectures. Electronics, 8(3), 292. MDPI AG. http://dx.doi.org/10.3390/electronics8030292

Hastie, T., Tibshirani, R., Friedman, J., & Franklin, J. (2005). The elements of statistical learning: data mining, inference and prediction. *The Mathematical Intelligencer, 27*(2), 83-85. http://thuvien.thanglong.edu.vn:8081/dspace/bitstream/DHTL_123456789/4053/1/%5BSpringer%20Series%20in%20Statistics-1.pdf

Huynh, T. D., Stalla-Bourdillon, S. & Moreau, L. (2019). *Provenance-based Explanations for Automated Decisions : Final IAA Project Report.* King's College London. https://kclpure.kcl.ac.uk/portal/en/publications/provenancebased-explanations-forautomated-decisions(5b1426ce-d253-49fa-8390-4bb3abe65f54).html

Rudin, C. (2019). Stop explaining black box machine learning models for high stakes decisions and use interpretable models instead. *Nature Machine Intelligence, 1*(5), 206. https://www.nature.com/articles/s42256-019-0048-x

LeCun, Y., Bengio, Y., & Hinton, G. (2015). Deep learning. *Nature 521.* http://www.cs.toronto.edu/~hinton/absps/NatureDeepReview.pdf

Leslie, D. (2020). *Explaining Decisions Made with AI.* The Alan Turing Institute and ICO. http://dx.doi.org/10.2139/ssrn.4033308

Rumelhart, D., Hinton, G., & Williams, R. (1986). Learning representations by back-propagating errors. *Nature 323.* 533–536. https://doi.org/10.1038/323533a0

Schmidhuber, J. (2015). Deep learning in neural networks: An overview. *Neural Networks 61.* https://arxiv.org/pdf/1404.7828.pdf

Shalev-Shwartz, S., & Ben-David, S. (2014). *Understanding machine learning: From theory to algorithms.* Cambridge University Press.

Waldrop, M., M. (2019). What are the limits of deep learning? *Proceedings of the National Academy of Sciences of the United States of America 116*(4). https://www.pnas.org/content/116/4/1074

## Sociotechnical Background

Amoore, L. (2021). The deep border. *Political Geography.* Advanced online publication: https://doi.org/10.1016/j.polgeo.2021.102547

Ball, K. (2009). Exposure: Exploring the subject of surveillance. *Information, Communication & Society, 12*(5), 639-657. https://doi.org/10.1080/13691180802270386

Beer, D. (2017). The social power of algorithms. *Information, Communication & Society, 20*(1), 1-13. https://doi.org/10.1080/1369118X.2016.1216147

Floridi, L., Cowls, J., Beltrametti, M., Chatila, R., Chazerand, P., Dignum, V., Luetge, C., Madelin, R., Pagallo, U., Rossi, F., Schafer, B., Valcke, P., & Vayena, E. (2018). AI4People—an ethical framework for a good AI society: opportunities, risks, principles, and recommendations. *Minds and machines, 28,* 689-707. https://doi.org/10.1007/s11023-018-9482-5

Jasanoff, S. (2015). Future imperfect: Science, technology, and the imaginations of modernity. *Dreamscapes of modernity: Sociotechnical imaginaries and the fabrication of power,* 1-33.

Kellogg, K. C., Valentine, M. A., & Christin, A. (2020). Algorithms at work: The new contested terrain of control, *Academy of Management Annals, 14*(1). https://doi.org/10.5465/annals.2018.0174

Makarius, E. E., Mukherjee, D., Fox, J. D., & Fox, A. K. (2020). Rising with the machines: A sociotechnical framework for bringing artificial intelligence into the organization. *Journal of Business Research, 120,* 262-273. https://doi.org/10.1016/j.jbusres.2020.07.045

Mehrabi, N., Morstatter, F., Saxena, N., Lerman, K., & Galstyan, A. (2021). A survey on bias and fairness in machine learning. *ACM Computing Surveys (CSUR), 54*(6), 1-35. https://arxiv.org/pdf/1908.09635

Mohamed, S., Png, M. T., & Isaac, W. (2020). Decolonial AI: Decolonial theory as sociotechnical foresight in artificial intelligence. *Philosophy & Technology, 33,* 659-684. https://doi.org/10.1007/s13347-020-00405-8

Morozov, E. (2013). *To save everything, click here: The folly of technological solutionism.* Public Affairs.

O'Neil, C. (2016). *Weapons of math destruction; How big data increases inequality and threatens democracy.* Crown.

Selbst, A. D., Boyd, D., Friedler, S. A., Venkatasubramanian, S., & Vertesi, J. (2019). Fairness and abstraction in sociotechnical systems. *In Proceedings of the conference on fairness, accountability, and transparency* (pp. 59-68). https://doi.org/10.1145/3287560.3287598

Taddeo, M., & Floridi, L. (2018). How AI can be a force for good. *Science, 361*(6404), 751-752. https://doi.org/10.1126/science.aat5991

Williams, R., & Edge, D. (1996). The social shaping of technology. *Research policy, 25*(6), 865-899. https://doi.org/10.1016/0048-7333(96)00885-2

# Responsible Research and Innovation: CARE and Act Principles

Anderson, E. (1999). What is the point of equality? *Ethics, 109,* 287–337. https://doi.org/10.1086/233897

Andrejevic, M., & Selwyn, N. (2020). Facial recognition technology in schools: Critical questions and concerns. *Learning, Media and Technology, 45*(2), 115-128. https://doi.org/10.1080/17439884.2020.1686014

Andrews, L., Benbouzid, B., Brice, J., Bygrave, L.A., Demortain, D., Griffiths, A., Lodge, M., Mennicken, A. & Yeung, K. (2017). Algorithmic regulation. *CARR, London School of Economics and Political Science Discussion Paper,(85)*. http://www.lse.ac.uk/accounting/assets/CARR/documents/D-P/Disspaper85.pdf

Arnold, M., Bellamy, R. K., Hind, M., Houde, S., Mehta, S., Mojsilovic, A., Nair, R., Ramamurthy, K. N., Reimer, D., Olteanu, A., Tsay, J., & Varshney, K. R & Piorkowski, D. (2018). FactSheets: Increasing Trust in AI Services through Supplier's Declarations of Conformity. *ArXiv.* https://arxiv.org/abs/1808.07261

Ashmore, R., Calinescu, R., & Paterson, C. (2019). Assuring the machine learning lifecycle: Desiderata, methods, and challenges. *ACM Computing Surveys (CSUR), 54*(5), 1-39. https://doi.org/10.1145/3453444

Bender, E. M., & Friedman, B. (2018). Data statements for natural language processing: Toward mitigating system bias and enabling better science. *Transactions of the Association for Computational Linguistics, 6,* 587-604. https://doi.org/10.1162/tacl_a_00041

Calo, R. (2017). Artificial Intelligence policy: a primer and roadmap. *UCDL Review, 51,* 399. https://heinonline.org/HOL/LandingPage?handle=hein.journals/davlr51&div=18&id=&page=&t=1560015127

D'Agostino, M., & Durante, M. (2018). Introduction: The governance of algorithms. *Philosophy & Technology, 31*(4), 499–505. https://doi.org/10.1007/s13347-018-0337-z

Fourcade, M., & Gordon, J. (2020). Learning like a state: Statecraft in the digital age. Journal of Law and Political Economy, 1(1), 78-108, https://doi.org/10.5070/LP61150258

Gebru, T., Morgenstern, J., Vecchione, B., Vaughan, J. W., Wallach, H., Daumé III, H., & Crawford, K. (2018). Datasheets for datasets. *ArXiv.* https://arxiv.org/abs/1803.09010

Holland, S., Hosny, A., Newman, S., Joseph, J., & Chmielinski, K. (2018). The dataset nutrition label: A framework to drive higher data quality standards. *ArXiv.* https://arxiv.org/abs/1805.03677

Leslie, D. (2019). *Understanding artificial intelligence ethics and safety: A guide for the responsible design and implementation of AI systems in the public sector.* The Alan Turing Institute. https://doi.org/10.5281/zenodo.3240529

Morley, J., Floridi, L., Kinsey, L., & Elhalal, A. (2019). From what to how. An overview of AI ethics tools, methods and research to translate principles into practices. *arXiv:1905.06876.* https://arxiv.org/abs/1905.06876

Reisman, D., Schultz, J., Crawford, K., & Whittaker, M. (2018). *Algorithmic impact assessments: a practical framework for public agency accountability.* AI Now. https://ainowinstitute.org/aiareport2018.pdf

Sanchez-Monedero, J., Dencik, L., & Edwards, L. (2020). What does it mean to solve the problem of discrimination in hiring? Social, technical and legal perspectives from the UK on automated hiring systems. *ArXiv.191006144 Cs.* https://doi.org/10.1145/3351095.3372849.

Saurwein, F., Just, N., & Latzer, M. (2015). Governance of algorithms: options and limitations. *info, 17*(6), 3549. https://www.emeraldinsight.com/doi/abs/10.1108/info-05-2015-0025

Spaulding, N. (2020). Is human judgment necessary?: Artificial intelligence, algorithmic governance, and the law. In M. K. Dubber, F. Pasquale & S. Das (Eds.), *The Oxford handbook of ethics of AI* (pp.375-402). Oxford University Press.

Tutt, A. (2016). An FDA for algorithms. *Administrative Law Review, 69*(1), 83-123. http://dx.doi.org/10.2139/ssrn.2747994

Wachter, S., & Mittelstadt, B. D. (2018). A right to reasonable inferences: re-thinking data protection law in the age of Big Data and AI. Columbia Business Law Review. https://ssrn.com/abstract=3248829

## Activities

Boal, A. (2000). *Theater of the Oppressed.* Pluto press.

Brand, J., & Sander, I. (2020). Critical data literacy tools for advancing data justice: A guidebook. *Data Justice Lab.* https://datajusticelab.org/wp-content/uploads/2020/06/djl-data-literacy-guidebook.pdf

Gangadharan, S. P., Petty, T., Lewis, T., & Saba, M. (2018). *Digital defense playbook: Community power tools for reclaiming data.* Our Data Bodies. http://www.odbproject.org/wp-content/uploads/2019/03/ODB_DDP_HighRes_Single.pdf

Lane, G., Angus, A., & Murdoch, A. (2018). UnBias Fairness Toolkit. https://unbias.wp.horizon.ac.uk/2018/09/06/unbias-fairness-toolkit/

Payne, B. H. (2019). An ethics of artificial intelligence curriculum for middle school students. *MIT Media Lab Personal Robots Group.* https://thecenter.mit.edu/wp-content/uploads/2020/07/MIT-AI-Ethics-Education-Curriculum.pdf

To find out more about the AI Ethics and
Governance in Practice Programme please visit:

turing.ac.uk/ai-ethics-governance

The Alan Turing Institute